\documentclass[twocolumn,showpacs,amsmath,amssymb,superscriptaddress,prc]{revtex4-2}
\usepackage{dcolumn,braket,color,amsmath,footnote,hyperref,bm}
\usepackage{graphicx,bm,url,longtable}
\usepackage{tabularx,multirow,braket}
\usepackage{fancybox}
\usepackage{tikz}
\usetikzlibrary{matrix}

\usepackage{hyperref}
\hypersetup{colorlinks=True,urlcolor=blue,linkcolor=blue,citecolor=blue,filecolor=black}

\colorlet{Changes@Color}{red}  % changes in red color

\newcommand\+{\dagger}

\newcommand\btm{\beta_{\mathrm{min}}}
\newcommand\gm{\gamma_{\mathrm{min}}}
\newcommand\am{\alpha_{\mathrm{min}}}
\newcommand\tbtm{\tilde\beta_{\mathrm{min}}}
\newcommand\tam{\tilde\alpha_{\mathrm{min}}}

\begin{document}

\title{
Interplay between pairing and triaxial shape degrees of freedom in Os and Pt nuclei
}

\author{K.~Nomura}
\affiliation{Department of Physics, Faculty of Science, 
University of Zagreb, HR-10000 Zagreb, Croatia}
\email{knomura@phy.hr}
\author{D.~Vretenar}
\affiliation{Department of Physics, Faculty of Science, University of
Zagreb, HR-10000 Zagreb, Croatia}
 \affiliation{ State Key Laboratory of Nuclear Physics and Technology, School of Physics, Peking University, Beijing 100871, China}
\author{Z.~P.~Li}
\affiliation{School of Physical Science and Technology, Southwest
University, Chongqing 400715, China}

\author{J.~Xiang}
\affiliation{School of Physics and Electronic, Qiannan Normal University
for Nationalities, Duyun 558000, China}
\affiliation{School of Physical Science and Technology, Southwest
University, Chongqing 400715, China}

\date{\today}

\begin{abstract}
The effect of coupling between pairing and quadrupole 
triaxial shape vibrations on the 
low-energy collective states 
of $\gamma$-soft nuclei is investigated using a model 
based on the framework of nuclear 
energy density functionals (EDFs).
Employing a constrained self-consistent mean-field (SCMF)
method that uses universal EDFs and pairing interactions,  
potential energy surfaces of characteristic $\gamma$-soft 
Os and Pt nuclei with $A\approx190$ are calculated as functions 
of the pairing and triaxial quadrupole deformations. 
Collective spectroscopic properties 
are computed using a number-nonconserving 
interacting boson model (IBM) Hamiltonian, with  
parameters determined by mapping the SCMF energy surface 
onto the expectation value of the 
Hamiltonian in the boson condensate state. 
It is shown that, by simultaneously considering both the shape 
and pairing collective degrees of freedom, the EDF-based IBM  
successfully reproduces data on collective structures based on low-energy 
$0^{+}$ states, as well as $\gamma$-vibrational bands. 
\end{abstract}

\maketitle

\section{Introduction}
Ground-state deformations of 
most medium-mass and heavy nuclei 
are of quadrupole type, parametrized by 
the axially-symmetric deformation $\beta$ 
(elongation along the 
symmetry axis of the intrinsic frame), 
and the degree of triaxiality $\gamma$ 
\cite{BM_II}. 
Quadrupole collectivity is a prominent 
feature of nuclei, and gives rise to  
interesting structure phenomena 
that include 
(i) quantum (shape) phase transitions 
\cite{cejnar2010} 
that correspond to sudden changes from nearly spherical vibrational 
to well-deformed rotational nuclear systems  
by addition or subtraction of just a few nucleons, and 
(ii) shape coexistence \cite{heyde2011} of two or 
more intrinsic shapes near the ground 
state in a single nucleus. 
For an accurate theoretical description 
of such phenomena, 
non-axial $\gamma$ deformations plays 
a crucial role. 
In this context, numerous theoretical 
studies have been carried out 
from various perspectives 
\cite{BM_II,RS,IBM,bender2003,caurier2005,cejnar2010,heyde2011,niksic2011,shimizu2012}.

In addition to shape degrees of freedom, 
pairing vibrations play an important role for 
the structure of heavy nuclei: In particular, 
for spectroscopic properties of excited 
$0^{+}$ states and the bands built on them, 
and for electric monopole ($E0$) transitions 
\cite{BM_II,bes1966,bes1970,bes1972,brink-broglia,garrett2016}. 
The relevance of the 
dynamical pairing degree of freedom in nuclear 
structure has been recognized 
since the development of the BCS theory for nuclei 
in the early 1960s 
\cite{bes1963,bes1966b,casten1972,ragnarsson1976}. 
The effect of dynamical 
pairing and its coupling 
to the (triaxial) quadrupole shape degrees of 
freedom has been studied using schematic models 
(see, e.g., \cite{prochniak1999,srebrny2006,prochniak2007}). 
Microscopic models, including those based 
on the self-consistent mean-field (SCMF) approaches, 
have also been employed to study 
the effects of pairing vibrations in various 
low-energy nuclear structure phenomena and 
fundamental nuclear processes such as the 
neutrinoless double-$\beta$ decay, 
\cite{vaquero2013} and spontaneous fission 
\cite{giuliani2014,zhao2016,rayner2018,rayner2020}. 
All these microscopic studies have been, 
however, limited to axially symmetric shapes, that is, 
calculations were performed within a two-dimensional 
(2D), pairing-plus-axial-quadrupole 
deformation space. 
A simultaneous quantitative treatment of 
pairing and triaxial quadrupole shape 
deformations, and their explicit coupling 
in realistic applications, has remained an unsolved 
problem for nearly sixty years.

In two recent studies we have introduced 
pairing vibrations as additional 
building blocks in the quadrupole collective model \cite{xiang2020}, 
and the interacting boson model (IBM) 
\cite{nomura2020pv}, 
based on the framework of 
nuclear energy density functionals (EDF). 
It was shown that the inclusion of the dynamical pairing 
significantly lowers the energies of excited $0^{+}$ states in 
deformed rare-earth nuclei \cite{xiang2020,nomura2020pv}. 
In Ref.~\cite{nomura2021pv} 
we have extended this framework 
to include both the triaxial 
quadrupole shape vibrations and pairing vibrations 
within the IBM. 
The method consists of two essential procedures.
Firstly, constrained SCMF
calculations have been performed using the 
relativistic mean-field plus BCS (RMF+BCS) 
method \cite{xiang2012} based on the 
PC-PK1 energy density functional \cite{PCPK1}, 
to construct the potential energy surface (PES) 
as a function of the three-dimensional (3D) 
quadrupole triaxial 
and pairing deformations (hereafter denoted 
as SCMF-PES). 
In a second step that takes into account both pairing 
vibrations and triaxiality in spectroscopic calculations, 
a boson-number-nonconserving 
Hamiltonian consisting of up to three-body 
boson terms has been introduced. 
The parameters of the IBM Hamiltonian are determined 
in such a way that the SCMF-PES in the vicinity of the 
global minimum in the 3D-deformation space is mapped 
onto the expectation value of the Hamiltonian 
in the boson condensate state (hereafter called IBM-PES). 
The diagonalization of the mapped IBM 
Hamiltonian in the Hilbert space that consists 
of three subspaces that differ in boson 
number by one,  
produces excitation spectra and transition rates. 
The method has been demonstrated in an  
illustrative application to the  
$\gamma$-soft nuclei $^{128}$Xe and $^{130}$Xe.

The aim of this work is to investigate in 
more detail 
the influence of simultaneously including 
the pairing and triaxial 
quadrupole shape vibrations on spectroscopic 
properties of $\gamma$-soft nuclei. 
Specifically, we give the formulation of the 
IBM framework within the 3D-deformation space 
and, in addition to the two cases  
already considered in Ref.~\cite{nomura2021pv} 
(i.e., $^{128,130}$Xe), 
extend the analysis to the mass 
$A\approx190$ nuclei: $^{188,190,192}$Os 
and $^{192,194,196}$Pt. 
The latter is another representative region  
in which non-axial deformations
play an important role. 
Previous studies within the IBM, based on the relativistic DD-PC1 \cite{DDPC1} EDF 
\cite{nomura2011coll,nomura2012tri}
and the Gogny-D1S \cite{D1S} and D1M \cite{D1M} 
EDFs \cite{nomura2011pt,nomura2011wos,nomura2011sys}, 
as well as with the five-dimensional collective 
Hamiltonian based on the relativistic PC-PK1 
\cite{yang2021}, have shown 
the importance of triaxiality in the mass $A\approx190$ region. 
An early empirical study  \cite{casten1978} presented
evidence for a rotor-to-O(6) 
transition in the Os-Pt region 
and, in particular, 
the $^{196}$Pt nucleus was shown to exhibit 
spectral features predicted in the O(6) 
dynamical symmetry limit of the IBM \cite{cizewski1978}. 
In addition to the PC-PK1 density functional, 
employed in Refs.~\cite{nomura2020pv,nomura2021pv}, 
here we also consider the DD-PC1 functional 
for the calculations of the Os and Pt isotopes. 
By comparing the results obtained with 
two representative EDFs, we examine 
the robustness of our method.

The paper is organized as follows. 
In Sec.~\ref{sec:method} 
we outline the theoretical method employed 
in the analysis of spectroscopic properties. The SCMF-PESs
and IBM-PESs in the 3D-deformation 
space are discussed in Sec.~\ref{sec:pes}.  
In Sec.~\ref{sec:spec} we analyze the calculated 
spectroscopic properties for the 
nuclei $^{128,130}$Xe, $^{188,190,192}$Os 
and $^{192,194,196}$Pt, including low-energy 
excitation spectra, the effect of coupling 
pairing and triaxial deformations 
on excited $0^{+}$ states and 
$\gamma$-vibrational bands, a comparison 
between the PC-PK1 and DD-PC1 functionals, 
and the $E2$ and $E0$ transition rates. Section~\ref{sec:conclusion}
contains a summary of the main results and 
an outline of future research.

\section{Method\label{sec:method}}
The SCMF-PESs are computed as functions of 
the 3D deformations by 
using the RMF+BCS method \cite{xiang2012}, 
with constraints on the mass quadrupole moments and intrinsic pairing deformation. 
The expectation values of the quadrupole operators 
$\hat{Q}_{20} = 2z^2 - x^2 - y^2$ and
$\hat{Q}_{22} = x^2 - y^2$ define the 
dimensionless polar deformation 
parameters $\beta$ and $\gamma$: 
\begin{align}
\label{eq:beta-f}
& \beta=\sqrt{\frac{5}{16\pi}}\frac{4\pi}{3}\frac{1}{A(r_{0}A^{1/3})^{2}}
\sqrt{\braket{\hat{Q}_{20}}^{2}+2\braket{\hat{Q}_{22}}^{2}} \\
& \gamma=\arctan{\sqrt{2}\frac{\braket{\hat{Q}_{22}}}{\braket{\hat{Q}_{20}}}}
\end{align}
with $r_0=1.2$ fm.
The expectation value of the 
monopole pairing operator 
$\hat{P}=({1}/{2}) \sum_{k>0}(c_{k}c_{\bar{k}}+c^\+_{\bar k}c^\+_{k})$  
in a BCS state (without pairing rotation), 
where $k$ and ${\bar{k}}$ denote the single-nucleon and
the corresponding time-reversed states, respectively, 
defines the intrinsic pairing deformation parameter $\alpha$:
\begin{align}
 \alpha
=\sum_{\tau=\pi,\nu}
\sum_{k>0}u_{k}^{\tau}v_{k}^{\tau} \;,
\end{align}
which can be related to the pairing gap $\Delta$.
To reduce the computational complexity, 
no distinction is made between proton and neutron 
pairing degrees of freedom even though, 
in principle, they should be treated separately.  
The particle-hole interactions are modeled 
by the relativistic energy density functionals  
PC-PK1 \cite{PCPK1} and DD-PC1 \cite{DDPC1}. 
For the particle-particle channel, a 
separable pairing force of finite range 
\cite{tian2009} is used.

Having computed the 3D SCMF-PES for 
a given nucleus, in the next step the deformation energy surface 
is mapped onto the corresponding 
interacting-boson system 
\cite{nomura2008,nomura2010}, using the procedure 
described below. 
Here the boson system consists of 
the monopole $s$ and quadrupole $d$ bosons, 
which, from a microscopic 
point of view \cite{OAI,IBM}, are associated with 
the correlated $L=0^+$ and $2^+$ pairs of 
valence nucleons, respectively. 
To take into account pairing vibrations, the 
number of bosons $n$, which equals half the number 
of valence nucleons \cite{OAI}, is not
conserved, but is allowed to vary by one unit, 
$n=n_{0},n_{0}\pm1$. 
The boson Hilbert space is then expressed 
as a direct sum of the three subspaces 
comprising 
$n=n_0-1$, $n_0$, and $n_0+1$ $sd$ bosons
\begin{align}
\label{eq:space}
 (sd)^{n_0-1}\oplus(sd)^{n_0}\oplus(sd)^{n_0+1}. 
\end{align}
In the following, the three subspaces 
are simply denoted  
by $[n_{0}-1]$, $[n_{0}]$, and $[n_{0}+1]$. 
The corresponding IBM Hamiltonian consists of 
the boson-number conserving (or unperturbed) 
$\hat{H}_\mathrm{cons}$
and non-conserving $\hat{H}_\mathrm{non-cons}$ 
interactions:
\begin{align}
 \label{eq:ham}
\hat{H}=\hat{H}_\mathrm{cons} + \hat{H}_\mathrm{non-cons}. 
\end{align}
To describe structures based on triaxial mean-field minima, 
it has been shown \cite{vanisacker1981,heyde1984,nomura2012tri} 
that it is necessary for the unperturbed 
Hamiltonian $\hat{H}_{\mathrm{cons}}$ to contain 
not only one- and two-body, 
but also three-body boson terms
\begin{align}
 \label{eq:cons}
\hat{H}_\mathrm{cons} = \hat{H}_{1b} + \hat{H}_{2b} + \hat{H}_{3b},
\end{align}
where
\begin{subequations}
 \begin{align}
\label{eq:h1b}
&\quad \hat{H}_{1b} = \epsilon_{d}\hat{n}_{d} + \hat{\delta} \\
\label{eq:h2b}
&\quad \hat{H}_{2b} = \kappa \hat{Q}\cdot\hat{Q} +
 \rho\hat{L}\cdot\hat{L} \\
\label{eq:h3b}
&\quad \hat{H}_{3b}
=
\eta\sum_{\lambda=2,4}
((d^\+ d^\+)^{(\lambda)} d^\+)^{(3)}\cdot ((\tilde d\tilde d)^{(\lambda)}\tilde d)^{(3)}
\end{align}
\end{subequations}
with the $d$-boson number operator 
$\hat {n}_d=\sum_{m}(-1)^{m}d^\+_{m}\cdot\tilde
d_{-m}$ ($\tilde d_{-m}=(-1)^{m}d_{m}$), 
the quadrupole operator
$\hat{Q}=s^\+\tilde d + d^\+s + \chi (d^\+\tilde d)^{(2)}$, 
and the angular momentum operator 
$\hat{L}=\sqrt{10}(d^\+\tilde d)^{(1)}$. 
The three-body boson interaction of the form 
(\ref{eq:h3b}) is shown to be particularly important 
to produce triaxial minima 
\cite{vanisacker1981,heyde1984,nomura2012tri}. 
We note that for three $d$ bosons there is only one state 
with angular momentum $L=3$ and, indeed, the terms with $\lambda=2$ 
and $\lambda=4$ in Eq.~(\ref{eq:h3b}) are proportional to 
each other. 
The term $\hat{\delta}=\epsilon_{0}\hat{n}
(=\epsilon_{0}(s^{\+}s+d^{\+}\cdot\tilde d))$ 
in $\hat{H}_\mathrm{1b}$ (\ref{eq:h1b})
determines the relative energies between the 
three unperturbed $0^+$ ground
states, but does not contribute 
to the excitation energies within 
each unperturbed boson subspace. 
The number-nonconserving Hamiltonian 
$\hat{H}_\mathrm{non-cons}$ (\ref{eq:ham}) is 
represented by a monopole-pair transfer operator
\begin{align}
\label{eq:noncons}
\hat{H}_\mathrm{non-cons}=\theta\frac{1}{2}(s^\+ + s), 
\end{align}
where $\theta$ denotes the strength parameter. 
The independent parameters of the total boson Hamiltonian 
(\ref{eq:ham}) are: $\epsilon_d$, $\kappa$, $\chi$, 
$\rho$, $\eta$, $\theta$, and $\epsilon_0$.

The IBM-PES within the 
$(\alpha,\beta,\gamma)$ 3D-deformation 
space is obtained by taking the expectation 
value of the Hamiltonian 
in the boson condensate 
state $\ket{\Psi(\vec\alpha)}$ 
\cite{ginocchio1980,dieperink1980,bohr1980}
\begin{align}
 \ket{\Psi({\vec\alpha})}=\ket{\Psi_{n_0-1}({\vec\alpha})}\oplus\ket{\Psi_{n_0}({\vec\alpha})}\oplus\ket{\Psi_{n_0+1}({\vec\alpha})}. 
\end{align}
The state $\ket{\Psi_{n}(\vec{\alpha})}$ 
for a given subspace comprising $n$ bosons 
($n=n_0-1,n_0,n_0+1$) is given by 
\begin{align}
\label{eq:coherent}
\ket{\Psi_{n}({\vec\alpha})}
=\frac{1}{\sqrt{n!}}(b^{\+}_{c})^{n}\ket{0},
\end{align}
where 
\begin{align}
%b^{\+}_{c}={\mathcal N}^{-1/2}\Biggl[\alpha_s
% s^\+ + \sum_{m=-2}^{+2}\alpha_{m}d_{m}^\+\Biggr]
b^{\+}_{c}=
\frac{1}{\sqrt{\mathcal{N}}}
\Biggl[
\alpha_{s}s^{\+} + \tilde{\beta}\cos{\gamma}d_{0}^{\+}
+\frac{1}{\sqrt{2}}\tilde{\beta}\sin{\gamma}(d_{+2}^{\+}+d_{-2}^{\+})
\Biggr]
\end{align}
with a normalization factor 
${\mathcal{N}}=\alpha_{s}^2+\tilde{\beta}^{2}$. 
The vector $\vec{\alpha}$ represents the 
three amplitudes 
$\{\alpha_{s},\tilde{\beta},\gamma\}$. 
$\ket{0}$ is the boson vacuum, that is, the inert core. 
The IBM-PES 
is expressed \cite{frank2004} as a $3\times 3$ matrix
${\bf E}(\vec{\alpha})$, with 
\begin{subequations}
 \begin{align}
\label{eq:pes-diag}
E_{n,n}(\vec{\alpha})
&=
\braket{\Psi_n(\vec{\alpha})|H_{\mathrm{cons}}|\Psi_n(\vec{\alpha})}
\nonumber \\
&=a_{0}n + n(a_1+a_2\tilde{\beta}^2){\mathcal{N}}^{-1} \nonumber \\
& +n(n-1)[
b_1\alpha_{s}^2\tilde{\beta}^2 + b_2\alpha_s\tilde{\beta}^3\Gamma
+ b_3\tilde{\beta}^4]{\mathcal{N}}^{-2}
\nonumber \\
& + d n(n-1)(n-2){\mathcal{N}}^{-3}\tilde{\beta}^6(1-\Gamma^2)
\end{align}
for the diagonal elements, and 
\begin{align}
\label{eq:pes-nondiag}
E_{n,n'}(\vec{\alpha})=E_{n',n}(\vec{\alpha})
&=\braket{\Psi_{n'}(\vec{\alpha})|H_{\mathrm{non-cons}}|\Psi_n(\vec{\alpha})}
\nonumber \\
&=\theta\alpha_s\sqrt{n+1}{\mathcal{N}}^{-1/2}
\end{align}
\end{subequations}
for the off-diagonal ones. 
Here the shorthand notations
$\Gamma\equiv\cos{3\gamma}$,
$a_0=\epsilon_0$, $a_1=5\kappa$, 
$a_2=\epsilon_d + 6\rho + \kappa(1+\chi^2)$,
$b_1=4\kappa$, 
$b_2=-4\sqrt{2/7}\kappa\chi$, 
$b_3=2\kappa\chi^2/7$,
and $d=-\eta/7$ are used. 
At each ($\alpha_{s},\tilde{\beta},\gamma$) 
coordinate, the matrix ${\bf E}(\vec{\alpha})$ 
is diagonalized, 
resulting in three energy surfaces. 
However, as it is often the case with  
configuration-mixing IBM calculations 
that deal with shape coexistence 
(e.g., Refs.~\cite{nomura2013hg,nomura2016zr,nomura2016sc,garciaramos2019}), 
only the lowest eigenvalue 
at each deformation is considered.

The amplitude $\tilde{\beta}$ is the IBM analog 
of the axially symmetric deformation $\beta$, 
while $\gamma$ represents the degree of triaxiality as usual. 
%identical to the geometrical counterpart. 
The following transformation of the variable 
$\alpha_{s}$ was introduced in \cite{nomura2020pv}: 
\begin{align}
\alpha_s = \cosh{(\tilde{\alpha}-\tilde{\alpha}_\mathrm{min})}. 
\end{align}
The new coordinate $\tilde{\alpha}$ is now considered the equivalent
quantity to the pairing deformation $\alpha$, 
and $\tilde{\alpha}_{\mathrm{min}}$ 
%and $\bar{\alpha}_{\mathrm{min}}\equiv c_{\alpha}\alpha_\mathrm{min}$
corresponds to the global minimum on the SCMF-PES. 
The $\tilde{\beta}$ and $\tilde{\alpha}$ 
variables in the boson system can be associated with 
the deformation parameters in the
SCMF model through the relations 
\cite{nomura2020pv,nomura2021pv}
\begin{align}
\label{eq:cab}
 \tilde{\alpha} = C_\alpha\alpha, \quad
 \tilde{\beta} = C_\beta\beta,
\end{align}
where the constants of proportionality 
$C_{\alpha}$ and $C_{\beta}$ are taken as additional 
parameters to be determined by the mapping. 
A well-known feature of the IBM is that 
the energy surface calculated in the condensate 
state is rather flat for large 
deformations far from the
global minimum, i.e., $\beta\gg\btm$
and $\alpha\gg\am$. 
This is a consequence of the fact
that the IBM is built on the restricted model space 
of valence nucleons, 
whereas the SCMF model considers all nucleons. 
This difference is partly taken into account 
by the rescaling relations in (\ref{eq:cab}). 
The scaling parameters $C_{\alpha}$ and $C_{\beta}$ 
should, in principle, be 
functions of deformations \cite{IBM}, 
hence we assume that they have the following 
$\alpha$ and $\beta$ dependencies. 
\begin{subequations}
\begin{align}
\label{eq:cb-new}
& C_{\beta}'
=C_{\beta}
[\theta(-\beta_{\ast})+\theta(\beta_{\ast})e^{p\beta_{\ast}^{2}}]
[\theta(-\alpha_{\ast})+\theta(\alpha_{\ast})e^{q\alpha_{\ast}^{2}}] \\
& C_{\alpha}'
\label{eq:ca-new}
=C_{\alpha}
[\theta(-\alpha_{\ast})+\theta(\alpha_{\ast})e^{r\alpha_{\ast}^{2}}]
\end{align}
\end{subequations}
with $\alpha_{\ast}=\tilde{\alpha}-{\tam}$ and
$\beta_{\ast}=\tilde{\beta}-{\tbtm}$, and 
the step function 
$\theta(x)(=1$, if $x\geqslant 0$ and $=0$, if $x<0$). 
The idea behind the above formulas 
(\ref{eq:cb-new}) and (\ref{eq:ca-new}) is that 
the IBM-PES can be made steeper for large 
deformations $\alpha\gg\am$ 
and $\beta\gg\btm$, 
so that it reproduces the SCMF deformation surface 
while, for relatively small deformations 
$\alpha\leqslant\am$ and 
$\beta\leqslant\btm$, the 
relations in (\ref{eq:cab}) hold, that is, 
$C'_{\alpha}$ and $C'_{\beta}$ are constant 
($C'_{\alpha}=C_{\alpha}$ and $C'_{\beta}=C_{\beta}$). 
Fixed values are used 
for the dimensionless coefficients:
$p=6$, $q=1$, and $r=0.1$.

The boson Hamiltonian (\ref{eq:ham}) 
is determined by applying the procedure 
of Refs.~\cite{nomura2020pv,nomura2021pv}. 
First, the parameters of the number-conserving
Hamiltonian (\ref{eq:cons})
$\{\epsilon_{d},\kappa,\chi,\eta,C_{\beta}\}$ 
are fixed by mapping the SCMF-PES 
in the 2D $(\beta,\gamma)$ space with 
$\alpha=\alpha_\mathrm{min}$ onto the 
diagonal matrix element 
associated with the normal $[n_{0}]$ 
configuration 
$E_{n_0,n_0}(\alpha_\mathrm{min},\beta,\gamma)$. 
Second, the strength parameter $\rho$ of the term 
$\hat{L}\cdot\hat{L}$ is determined  
separately \cite{nomura2011rot}, by equating 
the bosonic cranking moment of inertia in the
intrinsic frame at the global minimum
$(\am,\btm,\gm)$ to the 
corresponding Inglis-Belyaev 
(IB) value \cite{inglis1956,belyaev1961} 
computed using the SCMF quasiparticle states and energies. 
We note that the IB moment of inertia must 
be increased by 40 \% for Xe and Os nuclei 
both for the PC-PK1 and DD-PC1 EDFs. 
This is to take into account the well-known fact that 
the IB formula underestimates the empirical 
moments of inertia. 
Third, the number-nonconserving Hamiltonian 
(\ref{eq:noncons}) 
is determined in such a way that the 
SCMF-PES in the 2D $(\alpha,\beta)$ space 
with $\gamma=\gm$ is reproduced by the 
lowest eigenvalue 
of the matrix ${\bf E}(\vec{\alpha})$. 

Having thus determined the boson interaction parameters, 
the mapped IBM Hamiltonian (\ref{eq:ham}) is diagonalized 
in the model space defined by Eq.~(\ref{eq:space}).

%-----------------------------------------------------------
%	DFT PES -- PC-PK1, minimum
%-----------------------------------------------------------
\begin{figure*}[htb!]
\begin{center}
\includegraphics[width=\linewidth]{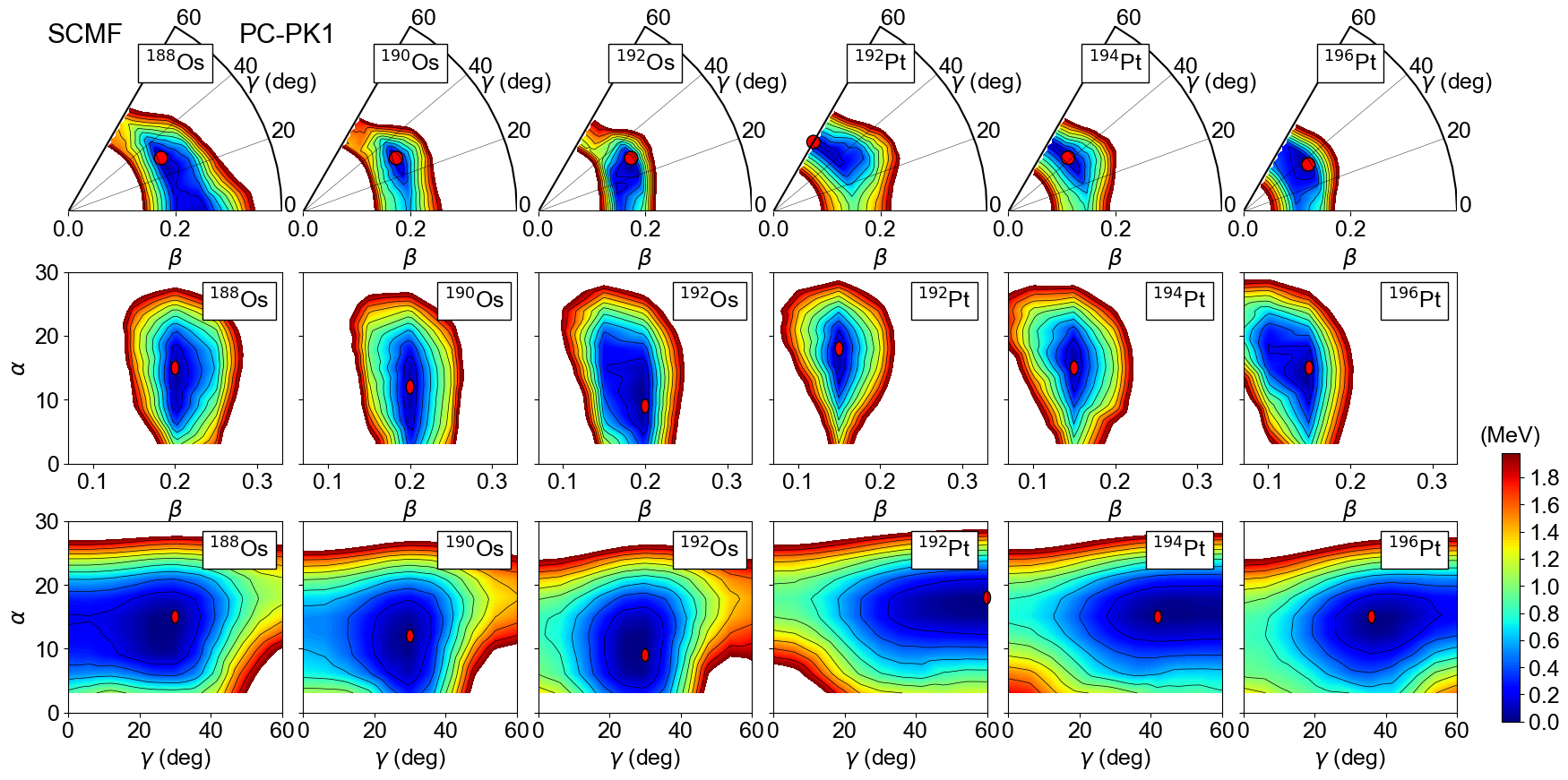}
\caption{The SCMF-PESs of  $^{188,190,192}$Os and $^{192,194,196}$Pt, 
projected onto the two-dimensional $(\beta,\gamma)$ [top], 
 $(\alpha,\beta)$ [middle], and  $(\gamma,\alpha)$ [bottom] 
planes (see text for the description). 
The fixed values of $\alpha$ in the $(\beta,\gamma)$, $\gamma$ in the 
$(\alpha,\beta)$, and $\beta$ in the $(\gamma,\alpha)$ plots, 
correspond to the global minimum in the entire three-dimensional 
$(\alpha,\beta,\gamma)$ PES (cf. Table \ref{tab:min}).
The PC-PK1 density functional and a separable pairing interaction have been  
used in the constrained SCMF calculation.}
\label{fig:pesmin_dft_pk}
\end{center}
\end{figure*}
%-----------------------------------------------------------
%	DFT PES -- DD-PC1, minimum
%-----------------------------------------------------------
\begin{figure*}[htb!]
\begin{center}
\includegraphics[width=\linewidth]{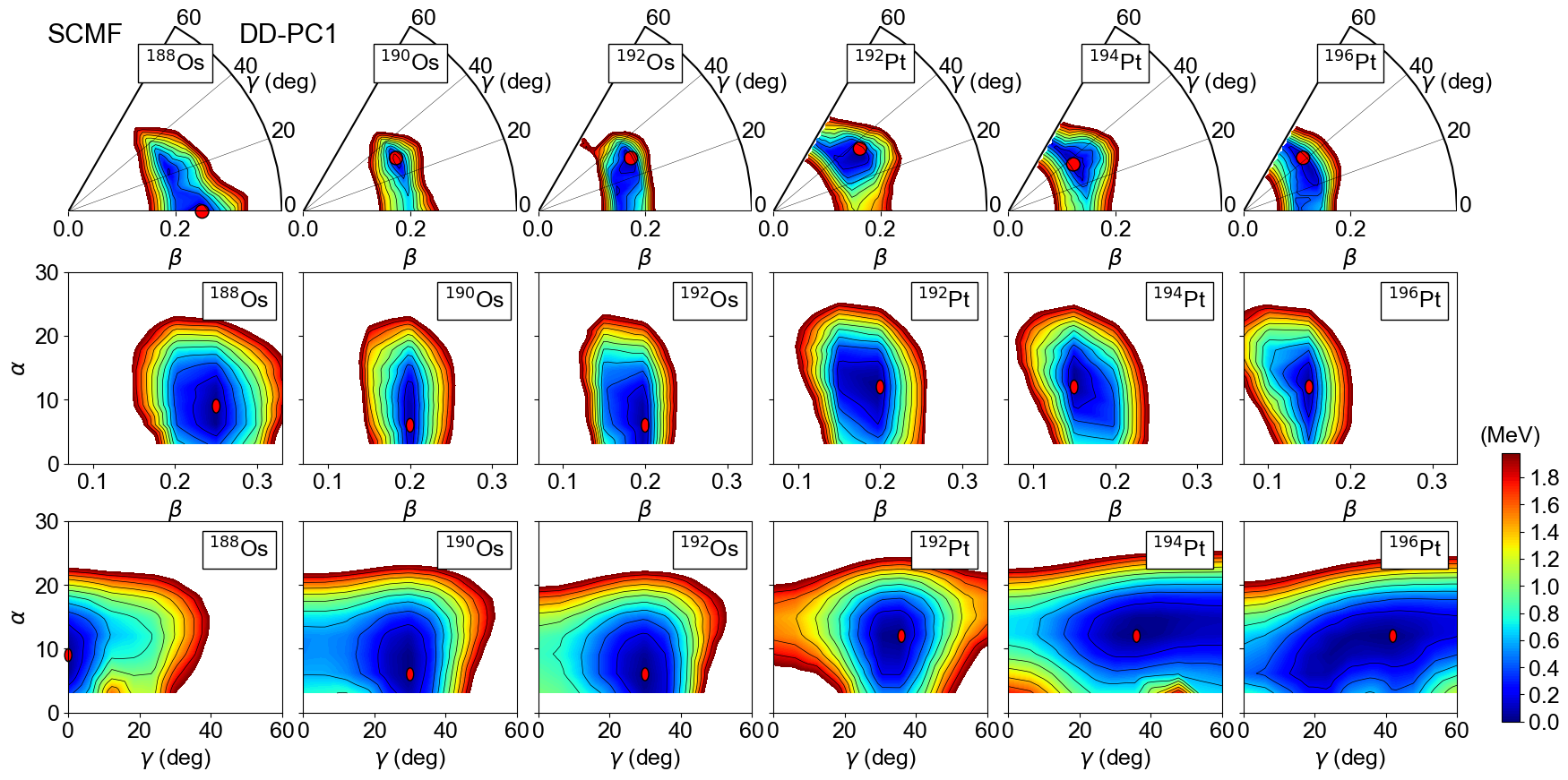}
\caption{Same as in the caption to Fig.~\ref{fig:pesmin_dft_pk}
but for the functional DD-PC1.}
\label{fig:pesmin_dft_dd}
\end{center}
\end{figure*}
%-----------------------------------------------------------
%	PESs -- Xe128, PC-PK1
%-----------------------------------------------------------
\begin{figure*}[htb!]
\begin{center}
\begin{tabular}{cc}
\includegraphics[width=0.5\linewidth]{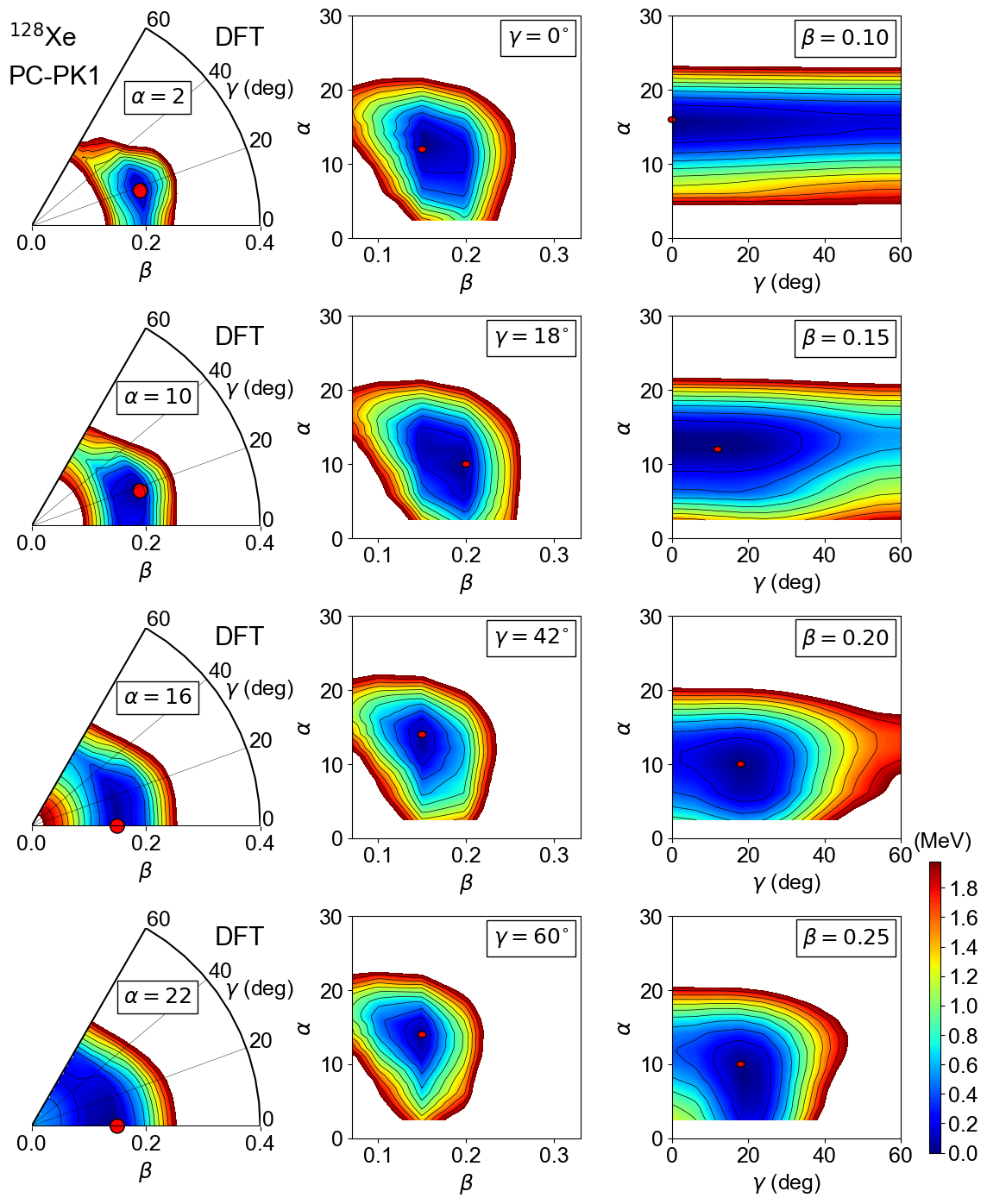}
 &
\includegraphics[width=0.5\linewidth]{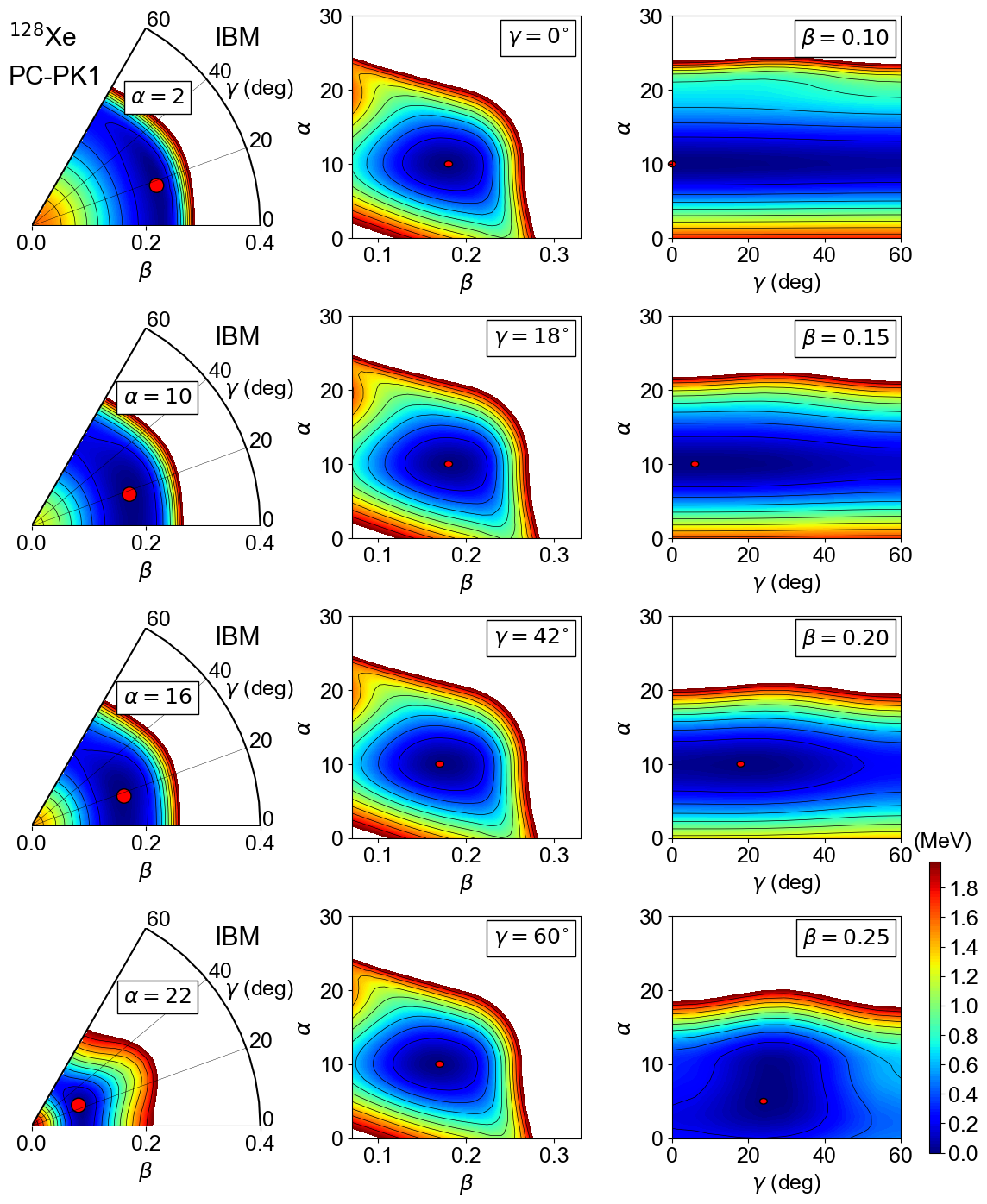}
\end{tabular}
\caption{The SCMF-PESs of $^{128}$Xe projected onto
the $(\beta,\gamma)$ [first column],
 $(\alpha,\beta)$ [second column],  
and $(\gamma,\alpha)$ planes [third column], as functions of the 
$\alpha$, $\gamma$, and $\beta$ 
deformations, respectively. 
The PC-PK1 density functional and a separable pairing interaction have been  
used in the constrained SCMF calculation.
The corresponding mapped IBM-PESs are plotted on the right hand side 
(columns $4 - 6$).
}
\label{fig:pes_xe128_pk}
\end{center}
\end{figure*}
%-----------------------------------------------------------
%	PESs -- Os188, PC-PK1
%-----------------------------------------------------------
\begin{figure*}[htb!]
\begin{center}
\begin{tabular}{cc}
 \includegraphics[width=0.5\linewidth]{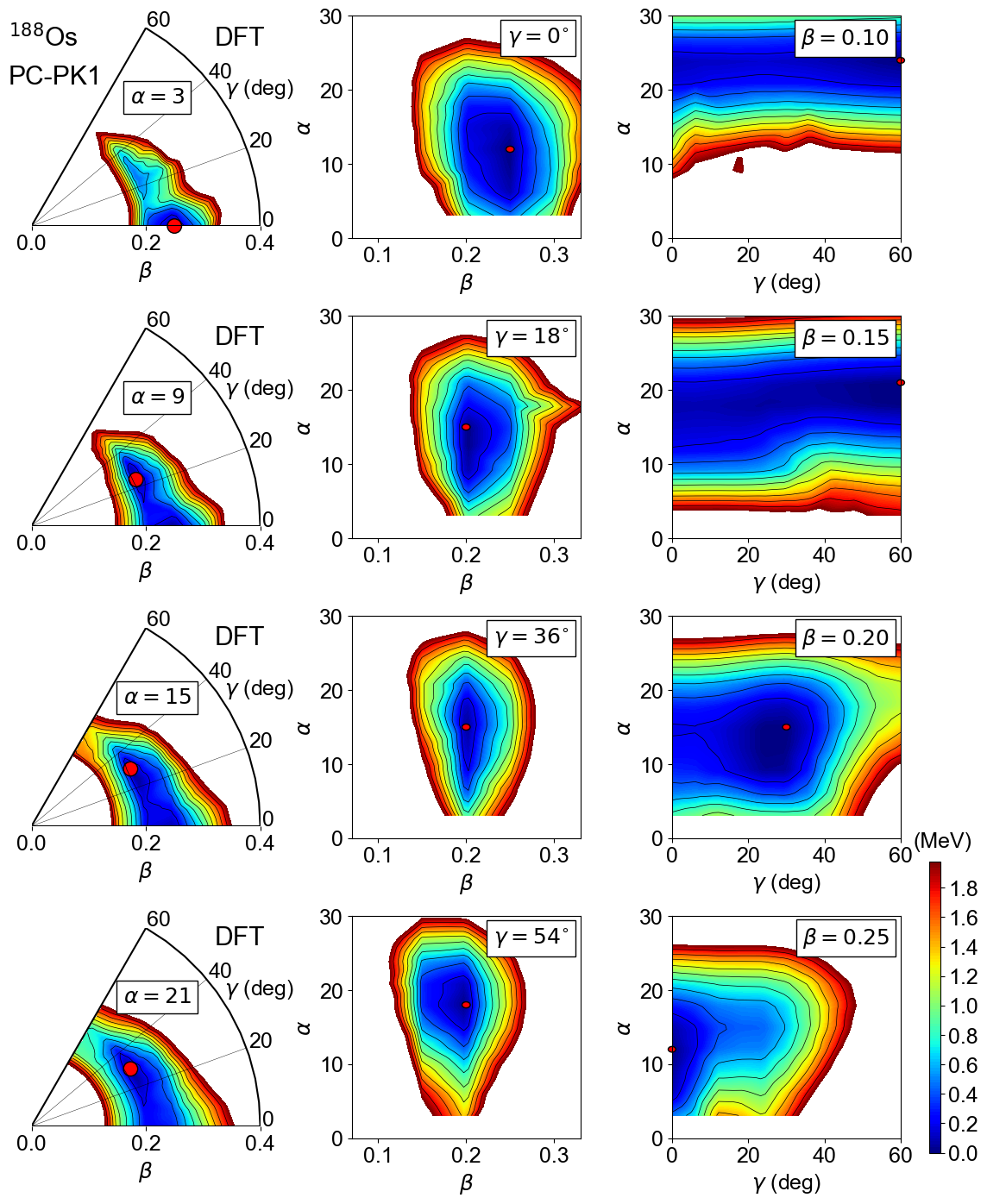}
 &
\includegraphics[width=0.5\linewidth]{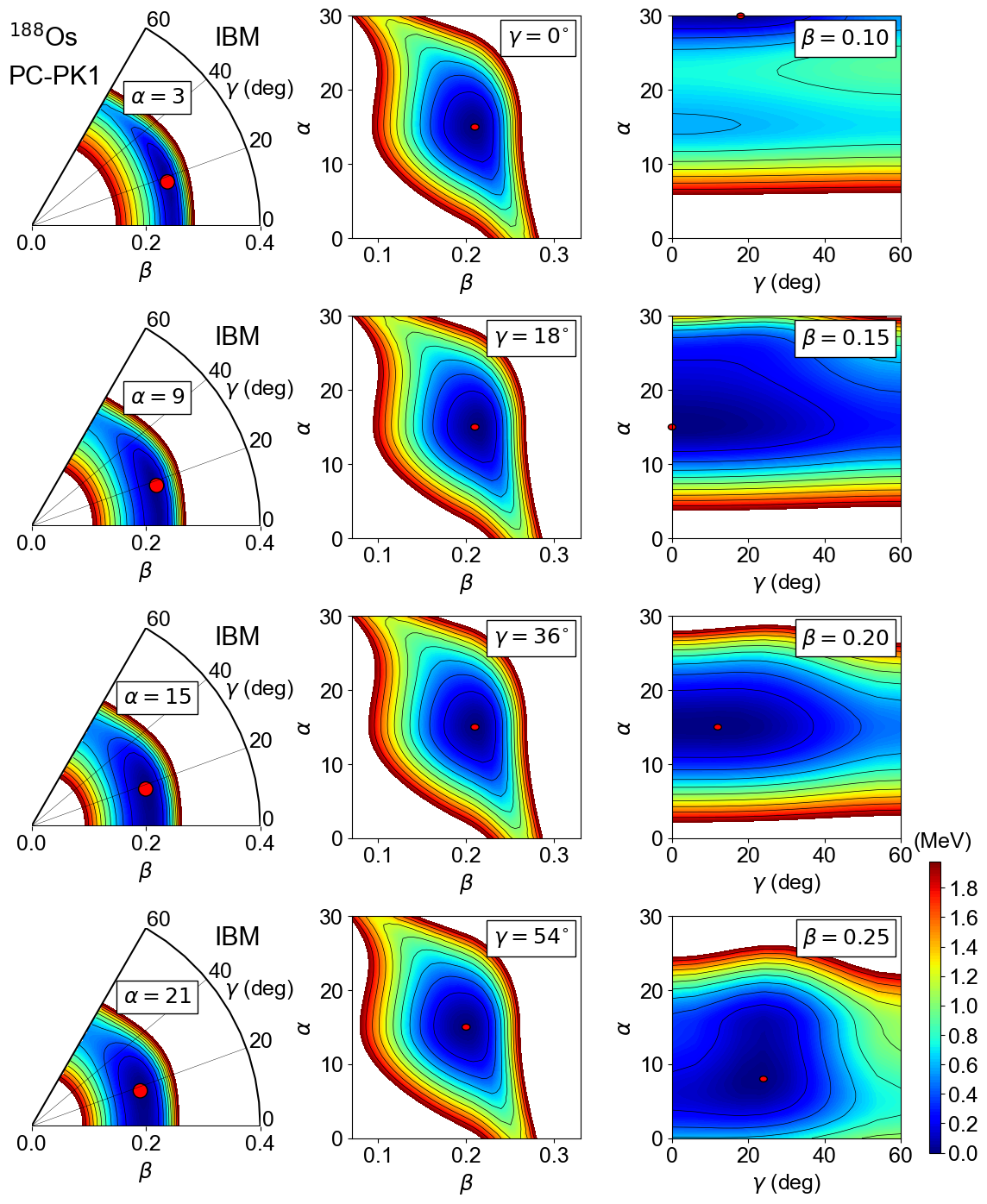}
\end{tabular}
\caption{Same as in the caption to Fig.~\ref{fig:pes_xe128_pk} but for $^{188}$Os.}
\label{fig:pes_os188_pk}
\end{center}
\end{figure*}
%-----------------------------------------------------------
%	PESs -- Pt194, PC-PK1
%-----------------------------------------------------------
\begin{figure*}[htb!]
\begin{center}
\begin{tabular}{cc}
\includegraphics[width=0.5\linewidth]{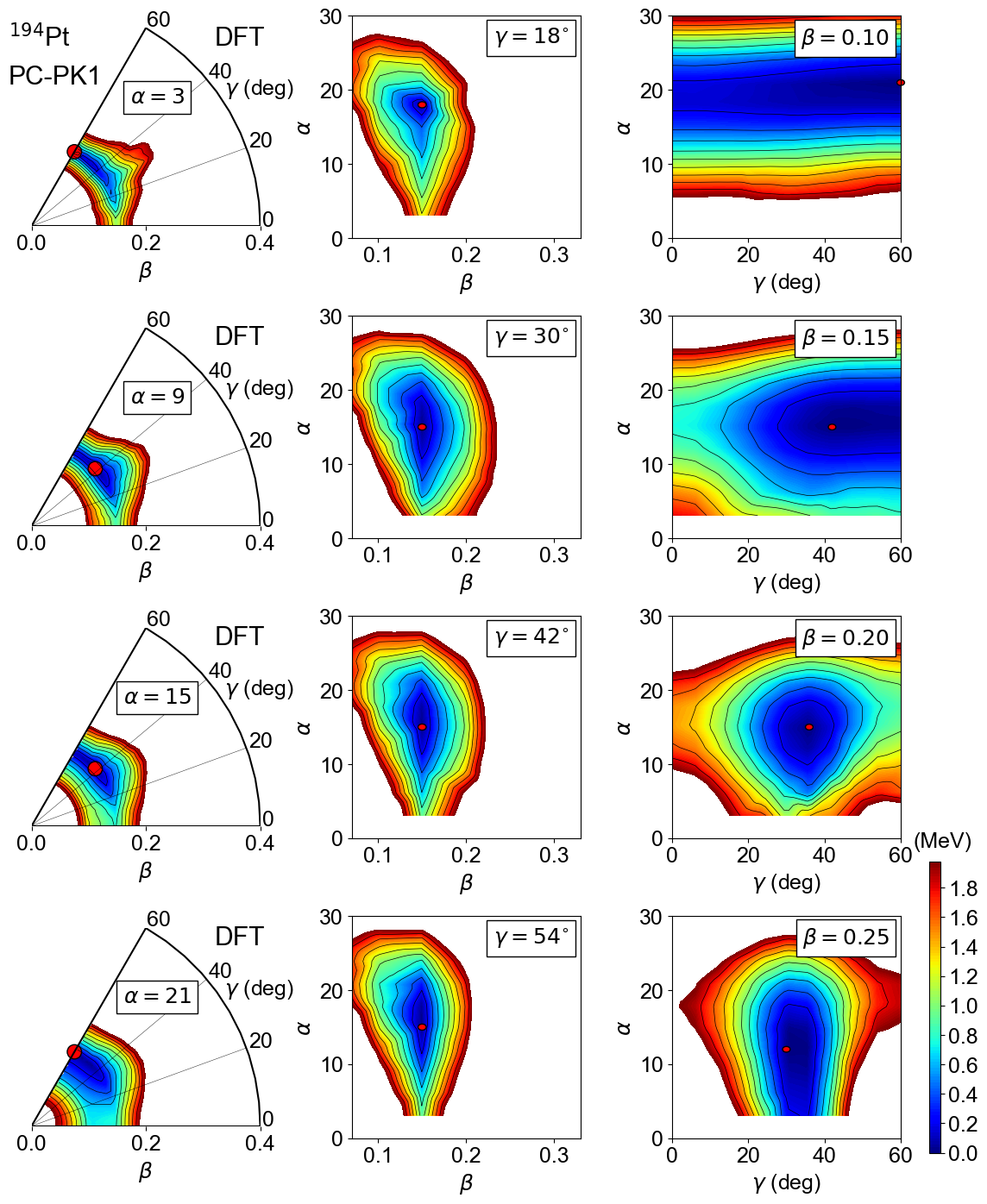}
 &
\includegraphics[width=0.5\linewidth]{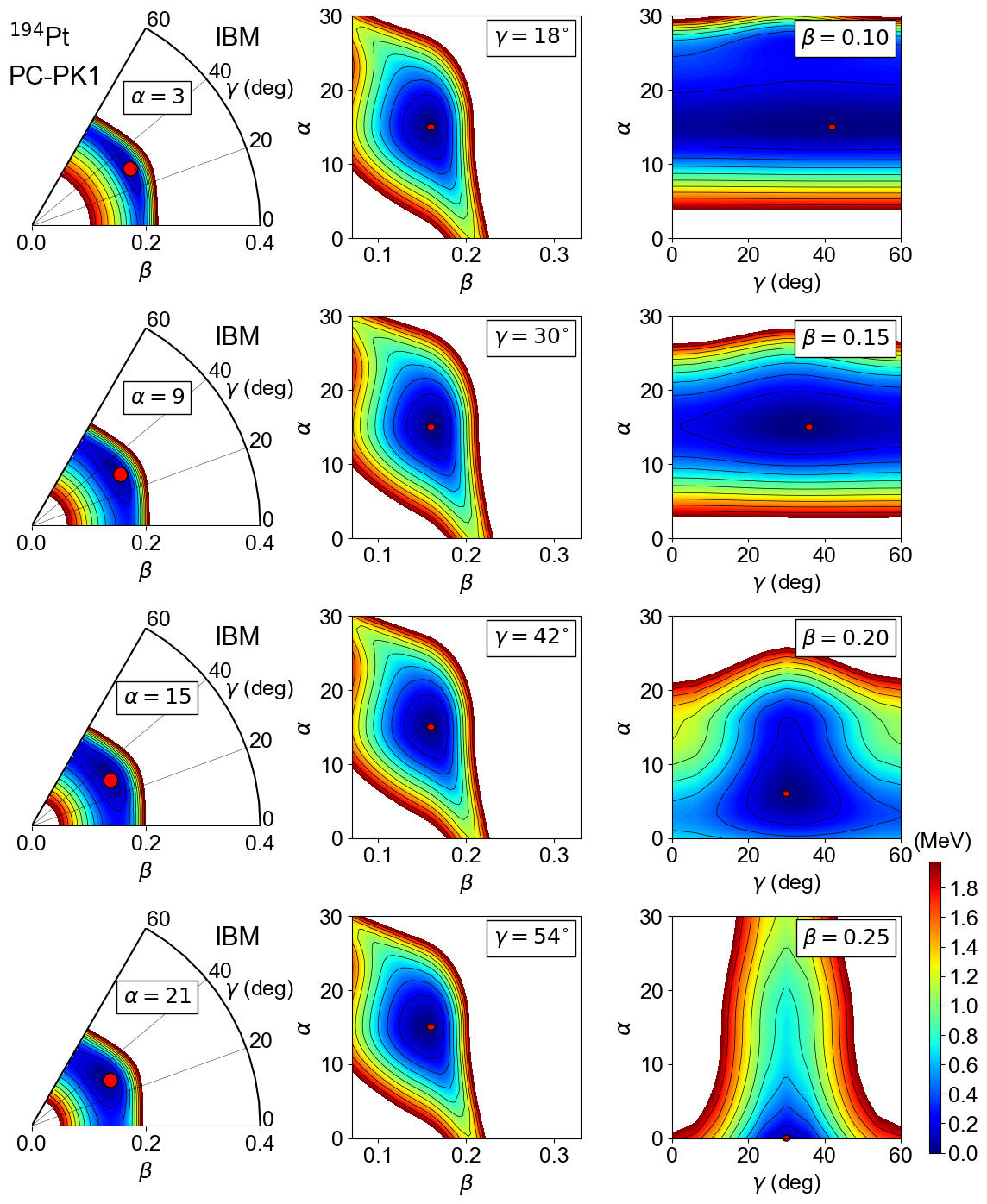}
\end{tabular}
\caption{Same as in the caption to Fig.~\ref{fig:pes_xe128_pk} but for $^{194}$Pt.}
\label{fig:pes_pt194_pk}
\end{center}
\end{figure*}

\section{Potential energy surfaces\label{sec:pes}}

\subsection{2D projections of SCMF-PESs}
The 3D energy surfaces for the Os and Pt isotopes, 
calculated using the self-consistent RMF+BCS model,  
are projected onto the 
2D $(\beta,\gamma)$, $(\alpha,\beta)$, 
and $(\gamma,\alpha)$
deformation spaces in Fig.~\ref{fig:pesmin_dft_pk} 
(PC-PK1) and Fig.~\ref{fig:pesmin_dft_dd} (DD-PC1), respectively. 
The 2D projections of the PESs are shown as functions 
of the axial quadrupole and
triaxial deformations $(\beta,\gamma)$ with fixed $\am$, 
axial quadrupole and pairing deformations 
$(\alpha,\beta)$ with fixed $\gm$, and triaxial quadrupole 
and pairing deformations ($\gamma,\alpha$) 
with fixed $\btm$. 
On each surface the fixed values of the deformation parameters  
correspond to the global minimum in the entire three-dimensional 
$(\alpha,\beta,\gamma)$ PES (cf. Table \ref{tab:min}).
Similar contour plots for the nuclei $^{128,130}$Xe 
can be found in Ref.~\cite{nomura2021pv}. 
The $\am$, $\btm$, and $\gm$ values for the 
considered nuclei are listed in Table~\ref{tab:min}. 
Below we mainly discuss prominent features of the SCMF-PESs
calculated with the energy density functional PC-PK1 
plus separable pairing, noting 
the very similar topology of the energy surfaces 
obtained using the DD-PC1 EDF.  

Considering first the $(\beta,\gamma)$ surfaces, 
shown in the top row of 
Fig.~\ref{fig:pesmin_dft_pk}, 
the isotopes $^{188,190}$Os appear to be 
$\gamma$-soft in the interval  
$0^{\circ}\leqslant\gamma\leqslant 40^{\circ}$ 
and rigid in $\beta$ deformation.
Remarkably, a rigid triaxial minimum 
is obtained near $\gamma=30^{\circ}$ 
for the nucleus $^{190}$Os. 
The three Pt nuclei also exhibit a degree of $\gamma$-softness, 
but are predominantly oblate in shape. 
The $(\alpha,\beta)$ energy surfaces, depicted 
in the middle row of Fig.~\ref{fig:pesmin_dft_pk}, 
are notably soft in $\alpha$ deformation, while
rather rigid in the axial $\beta$ deformation.  
The $\alpha$-softness indicates pronounced  
pairing fluctuations. The $\am$ values for the Pt nuclei 
are generally larger than those for the Os isotopes. 
In both Os and Pt nuclei, the equilibrium 
minimum $\am$ in the 
$(\alpha,\beta)$ plane gradually decreases 
with neutron number. 
In the bottom row of Fig.~\ref{fig:pesmin_dft_pk}, 
we note that the SCMF-PESs 
in the $(\gamma,\alpha)$ plane show more variation 
with nucleon number. As one can already infer 
from the two top rows, the $(\gamma,\alpha)$ 
surfaces are soft in both coordinates.
Generally, the Os nuclei are more soft in 
$\alpha$, while $\gamma$-softness on the oblate 
side ($\gamma\approx60^{\circ}$) 
is more pronounced in the Pt isotopes. 

At the quantitative level, there are 
certain differences in the topology of the 
SCMF-PESs obtained using the two functionals. 
First, in all the 2D-deformation subspaces, 
the SCMF-PESs calculated with the 
PC-PK1 EDF are generally  
softer than those corresponding to the 
DD-PC1 EDF. 
For the $(\alpha,\beta)$ energy surfaces, 
the PC-PK1 EDF predicts larger equilibrium 
values $\am$ of pairing deformation 
(see also, Table~\ref{tab:min}). A pronounced difference 
between the two EDFs is also seen in the 
$(\gamma,\alpha)$ plane for the Pt isotopes. 
It is of interest that 
the DD-PC1 $(\gamma,\alpha)$-PESs for the 
$^{194,196}$Pt nuclei are softer in 
$\gamma$ deformation than those obtained with 
the PC-PK1 EDF. For $^{192}$Pt, the 
PC-PK1 SCMF-PES is $\gamma$ soft, while 
the one calculated with the DD-PC1 EDF 
is soft in $\alpha$ and 
more rigid in $\gamma$. 
\begin{table}
\caption{\label{tab:min}
The values ($\am,\btm,\gm$) of the deformation parameters 
at which global minima occurs on the 3D energy surfaces, for 
constrained SCMF calculations based in the functionals PC-PK1 
and DD-PC1.}
 \begin{center}
 \begin{ruledtabular}
  \begin{tabular}{ccc}
 & PC-PK1 & DD-PC1 \\
\hline
$^{128}$Xe & (10, 0.20, 18$^{\circ}$) & \\
$^{130}$Xe & (12, 0.15,  0$^{\circ}$) & \\
$^{188}$Os & (15, 0.20, 30$^{\circ}$) & (9, 0.25, 0$^{\circ}$) \\
$^{190}$Os & (12, 0.20, 30$^{\circ}$) & (6, 0.20, 30$^{\circ}$) \\
$^{192}$Os & ( 9, 0.20, 30$^{\circ}$) & (6, 0.20, 30$^{\circ}$) \\
$^{192}$Pt & (18, 0.15, 60$^{\circ}$) & (12, 0.20, 36$^{\circ}$) \\
$^{194}$Pt & (15, 0.15, 42$^{\circ}$) & (12, 0.15, 36$^{\circ}$) \\
$^{196}$Pt & (15, 0.15, 36$^{\circ}$) & (12, 0.15, 42$^{\circ}$) \\
  \end{tabular}
 \end{ruledtabular}
 \end{center}
\end{table}

\subsection{2D SCMF-PESs as functions of the third collective coordinate}
It is also of interest to consider, 
for individual nuclei, 
the variations of the 2D SCMF-PESs as functions of 
the third deformation variable.
As illustrative examples, 
we plot in Figs.~\ref{fig:pes_xe128_pk}, 
\ref{fig:pes_os188_pk}, 
and \ref{fig:pes_pt194_pk} the SCMF-PESs for the 
nuclei $^{128}$Xe, $^{188}$Os, and $^{194}$Pt, 
respectively. 
Here only the results obtained with the functional PC-PK1 are 
shown. The DD-PC1 results are very similar. 

We note some general features of the SCMF-PESs.  
(i) The $(\beta,\gamma)$-PESs for each nucleus 
become considerably softer as the intrinsic 
pairing deformation $\alpha$ is increased. 
This is particularly pronounced for $^{128}$Xe  
(Fig.~\ref{fig:pes_xe128_pk}). 
(ii) The $(\alpha,\beta)$ SCMF-PESs 
are less sensitive to the 
variation of the $\gamma$ deformation, but appear 
markedly soft near the 
global minimum corresponding to 
$\gamma=\gm$. 
(iii) The topology of the 
$(\gamma,\alpha)$-SCMF-PESs 
varies most rapidly with increasing $\beta$ values. 
For small values of $\beta$, 
these 2D surfaces exhibit pronounced $\gamma$ softness, and  
become softer in $\alpha$
with increasing $\beta$ deformation. 
Especially near the $\beta=\btm$ equilibrium 
values, we note a transitional feature, 
that is, the surfaces 
are particularly soft with respect to both 
$\alpha$ and $\gamma$.

\subsection{Mapped IBM-PESs}
From the fourth to the sixth columns of 
Figs.~\ref{fig:pes_xe128_pk}, \ref{fig:pes_os188_pk}, 
and \ref{fig:pes_pt194_pk}, we also depict 
the corresponding IBM-PESs. 
The variations of these energy surfaces  
as functions of the parameters
$\alpha$, $\gamma$, and $\beta$ on 
each of the 2D-$(\beta,\gamma)$, $(\alpha,\beta)$, 
and $(\gamma,\alpha)$ deformation spaces, 
respectively, reproduce those observed for the SCMF-PESs. 
A notable difference between the SCMF- and IBM-PESs 
in each 2D space is that the latter are considerably softer 
with respect to $\beta$ and $\gamma$ deformations. 
As explained in the previous section, such a difference 
arises because of the restricted (valence) 
model space of the IBM, compared to that of the 
SCMF model. 
Another reason is, apparently, the 
limited analytical form of the IBM-PES 
(\ref{eq:pes-diag}) and  (\ref{eq:pes-nondiag}),
which does not provide enough variability to 
accurately reproduce the topology of the 
SCMF-PES. 
For example, for general three-body boson 
terms the energy surface can have a more complicated 
$\gamma$ dependence, consisting of 
terms proportional to 
$\tilde{\beta}^{3}\Gamma$, $\tilde{\beta}^{5}\Gamma$, 
and $\tilde{\beta}^{6}\Gamma^{2}$, with $\Gamma=\cos{3\gamma}$. 
These would imply additional parameters, 
and thus we use the specific three-body term
of the type (\ref{eq:h3b}). 
On each 2D $(\beta,\gamma)$ and $(\gamma,\alpha)$ 
surfaces, the difference between the SCMF- and 
IBM-PESs becomes more pronounced as one moves away 
from the global minimum, because the mapping is 
considered only in the vicinity of the global minimum. 
However, these differences should not significantly affect 
the calculated spectroscopic properties of low-energy collective 
states.

\begin{table*}
\caption{\label{tab:para} 
The values of the IBM Hamiltonian parameters, 
as well as the constants of proportionality 
(\ref{eq:cab}), for $^{128,130}$Xe, 
$^{188,190,192}$Os and
 $^{192,194,196}$Pt, determined by the mapping of the 
 PC-PK1 plus separable-pairing self-consistent mean-field results. 
 The corresponding parameters obtained with  
the DD-PC1 EDF are shown in parentheses.}
 \begin{center}
 \begin{ruledtabular}
  \begin{tabular}{ccccccccc}
 & $^{128}$Xe & $^{130}$Xe
 & $^{188}$Os & $^{190}$Os & $^{192}$Os
 & $^{192}$Pt & $^{194}$Pt & $^{196}$Pt \\
\hline
$\epsilon_{d}$ (keV) 
& 236.5 & 218.4 & 117.8 (178.7) & 127.4 (191.5) & 111.7 (156.8) & 200 (249.6) & 150 (249.1) & 150 (251.5) \\
$-\kappa$ (keV) 
& 102 & 102 & 65 (80) & 65 (80) & 65 (80) & 65 (80) & 65 (80) & 65 (80)\\
$\chi$ 
& $-0.08$ & $-0.14$ & $-0.1$ ($-0.2$) & $-0.02$ ($-0.025$) & $-0.02$ ($-0.05$) & 0.2 (0.03) & 0.03 (0.03) & 0.03 (0.08) \\
$-\rho$ (keV) 
& 15.1 & 16.1 & 6.3 (16.4) & 7.9 (15.3) & 5.3 (9.5) & 0.0 (8.3) & 0.0 (8.2) & 0.0 (8.6) \\
$\eta$ (keV) 
& 70 & 100 & 25 (30) & 25 (30) & 30 (40) & 35 (30) & 40 (45) & 40 (60)\\
$\epsilon^{0}$ (MeV) 
& 1.35 & 1.3 & 1.55 (1.95) & 1.4 (1.7) & 1.3 (1.55) & 1.25 (1.5) & 1.1 (1.35) & 1.0 (1.2)\\
$\theta$ (keV) 
& 540 & 740 & 360 (360) & 360 (360) & 360 (360) & 360 (360) & 360 (360) & 360 (360) \\
$C_{\beta}$ 
& 4.0 & 4.3 & 4.0 (3.9) & 4.1 (4.1) & 4.4 (4.4) & 5.2 (4.2) & 5.0 (4.6) & 4.8 (5.0) \\
$C_{\alpha}$ 
& 0.09 & 0.095 & 0.07 (0.07) & 0.05 (0.05) & 0.07 (0.05) & 0.085 (0.06) & 0.08 (0.05) & 0.09 (0.08) \\
  \end{tabular}
 \end{ruledtabular}
 \end{center}
\end{table*}
The parameters of the IBM Hamiltonian (\ref{eq:ham}), 
as well as the proportionality coefficients $C_{\beta}$ 
and $C_{\alpha}$, for the Xe, Os, and Pt 
nuclei, obtained using the
mapping procedure described in the previous section, 
are listed in Table~\ref{tab:para}. 
For both EDFs, most of the parameters appear to 
be constant or vary only gradually with 
neutron number in each isotopic chain. 
It is satisfying that the parameters 
are only weakly dependent on the nucleon number, 
because this indicates the consistency of 
the method and supports the model predictions. 
The very small values of the parameter $\chi$ in 
the quadrupole operator are 
characteristic for $\gamma$-soft nuclei. To simplify 
the calculation, the mixing strength $\theta$ is constant 
($\theta=360$ keV) for all Os and Pt nuclei.

%-----------------------------------------------------------
%	Spectra, PC-PK1
%-----------------------------------------------------------
\begin{figure*}[htb!]
\begin{center}
 \includegraphics[width=\linewidth]{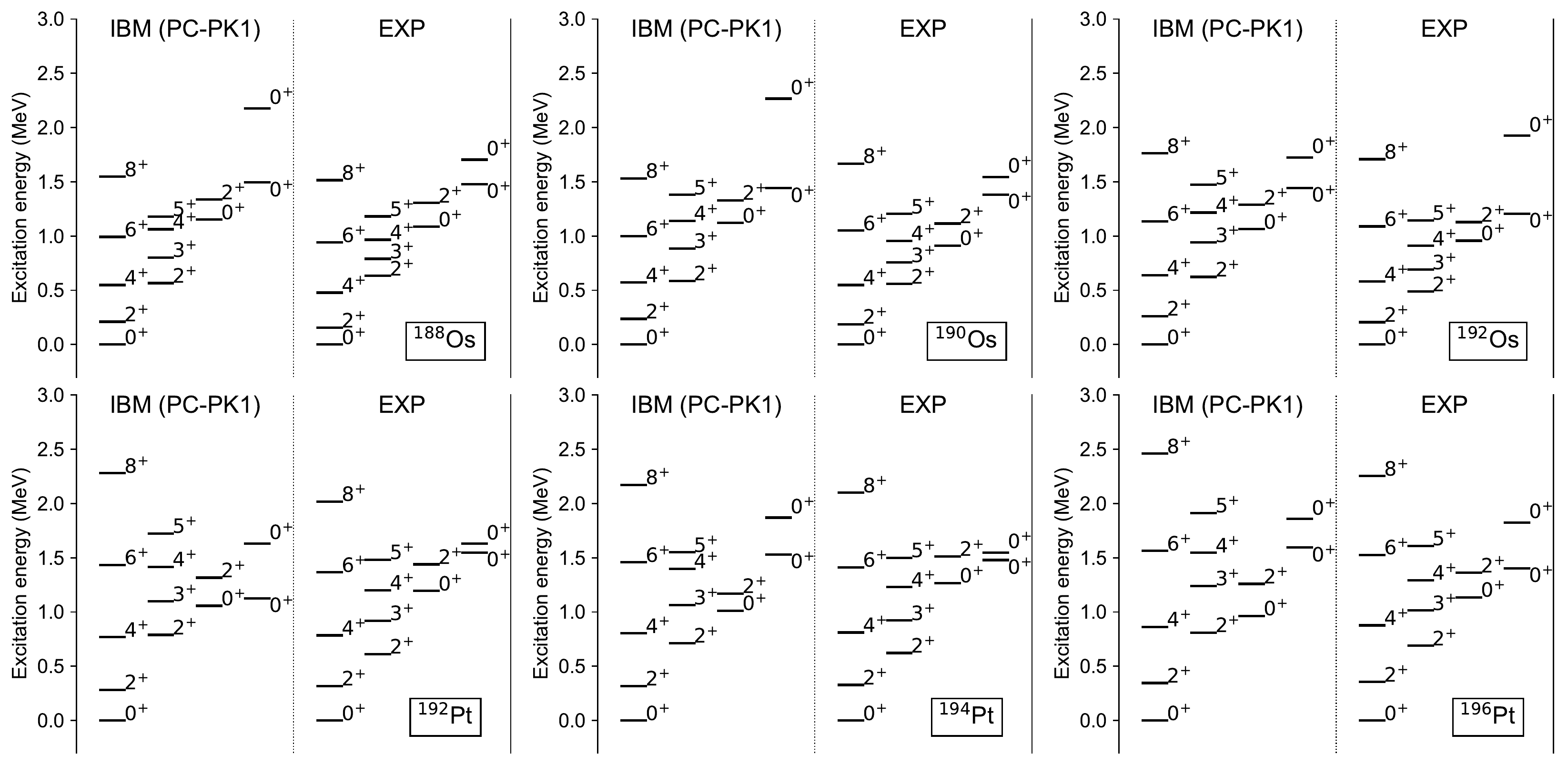}
\caption{Comparison of the 
experimental \cite{data} and theoretical 
excitation spectra of $^{188,190,192}$Os 
and $^{192,194,196}$Pt,  obtained using the IBM 
with the triaxial quadrupole 
$(\beta,\gamma)$ plus pairing $\alpha$ degrees of freedom,   
and based on the PC-PK1 microscopic energy density functional.} 
\label{fig:band_all_pk}
\end{center}
\end{figure*}
%-----------------------------------------------------------
%	Spectra, DD-PC1
%-----------------------------------------------------------
\begin{figure*}[htb!]
\begin{center}
 \includegraphics[width=\linewidth]{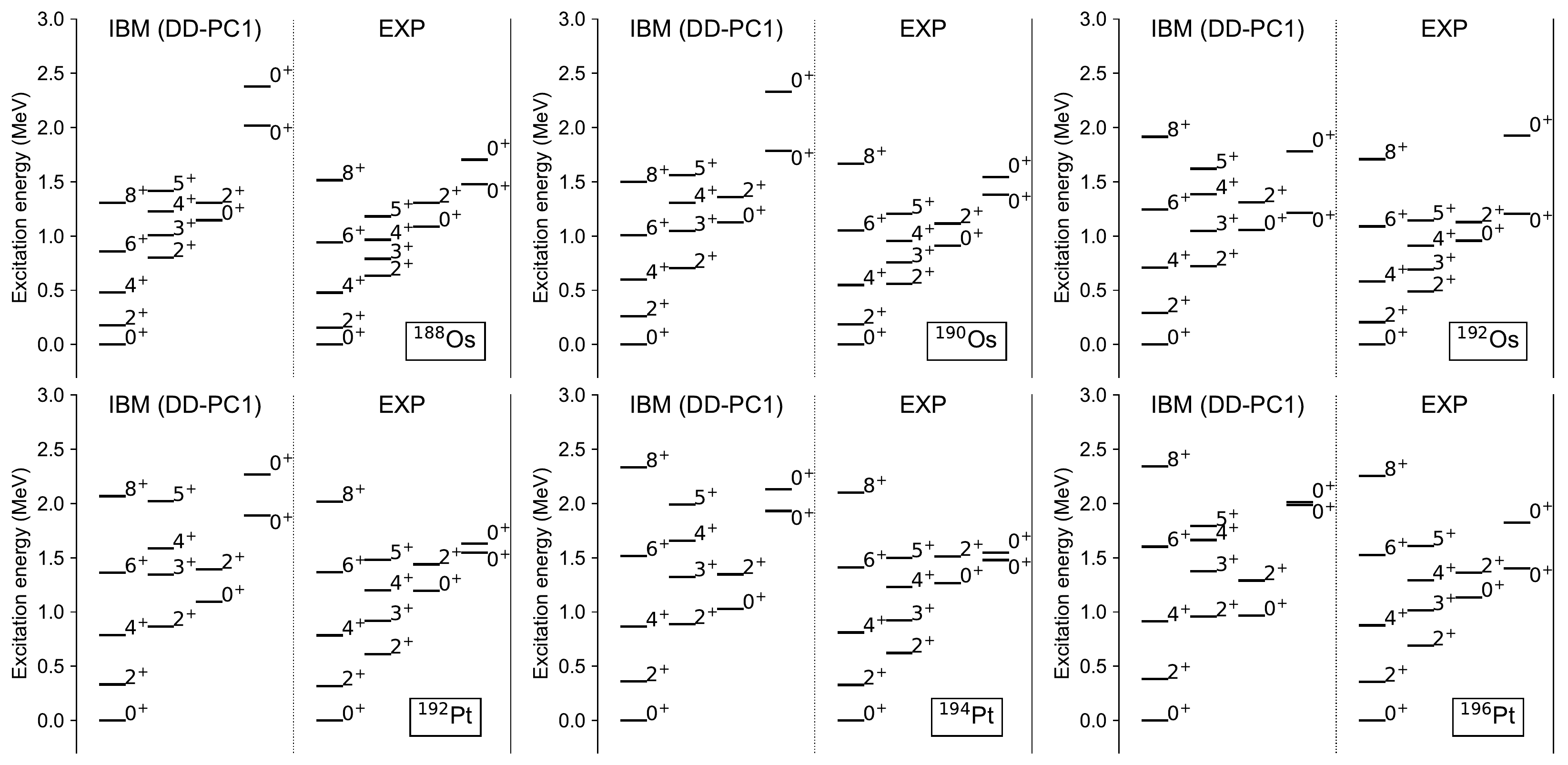}
\caption{Same as in the caption to Fig.~\ref{fig:band_all_pk}, 
but the DD-PC1 functional is used to determine the parameters of the IBM Hamiltonian} 
\label{fig:band_all_dd}
\end{center}
\end{figure*}

%-----------------------------------------------------------
%	0+ energies
%-----------------------------------------------------------
\begin{figure}[ht!]
\begin{center}
\includegraphics[width=\linewidth]{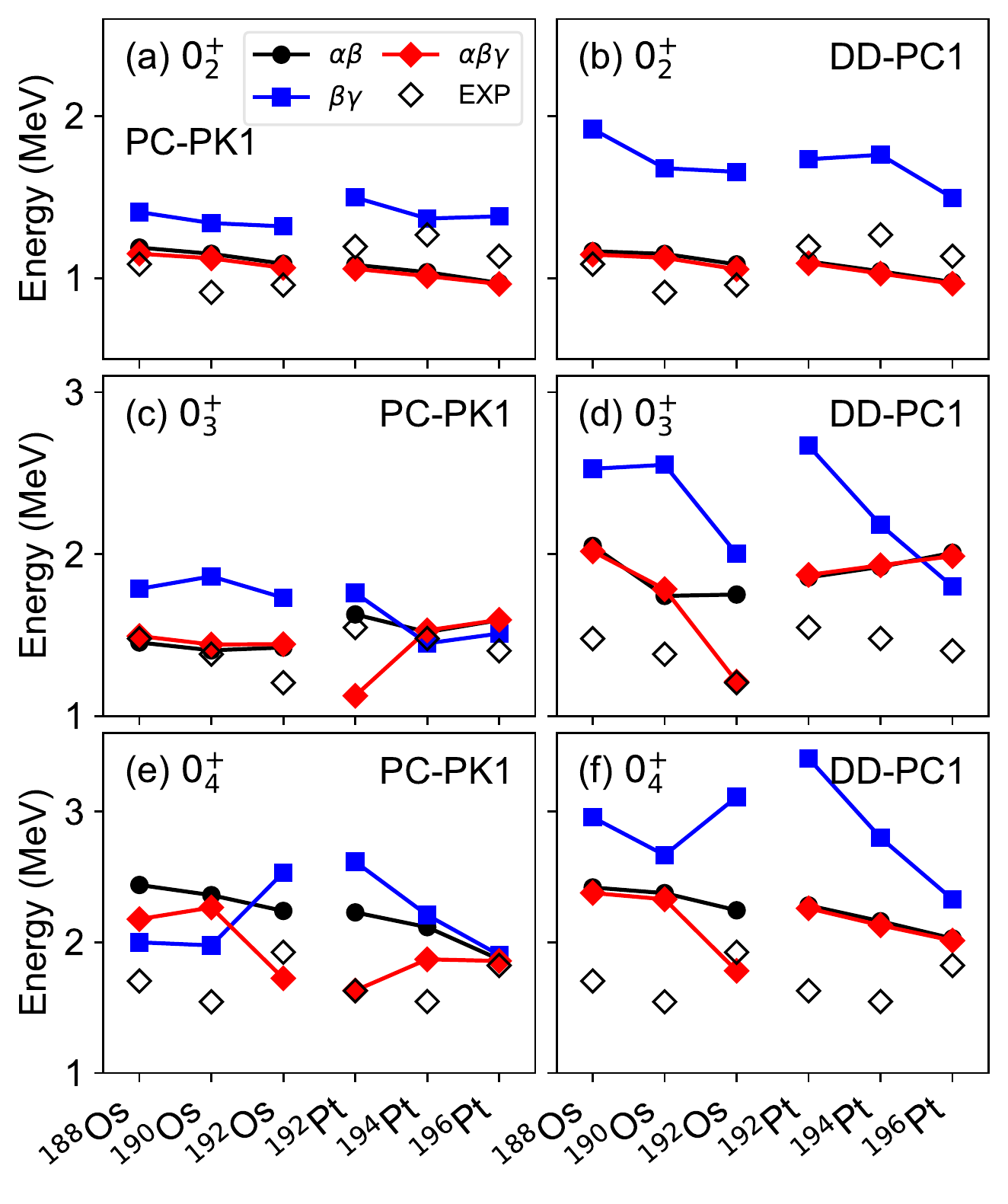}
\caption{Excitation energies of the three 
lowest $0^{+}$ excited states in $^{188,190,192}$Os 
and $^{192,194,196}$Pt, obtained from IBM
calculations that include the 2D (triaxial quadrupole $(\beta,\gamma)$), 2D
(axial quadrupole $\beta$ plus pairing $\alpha$), and 3D (triaxial
 quadrupole $(\beta,\gamma)$ and pairing
 $\alpha$) degrees of freedom, in comparison with the 
 corresponding experimental levels.} 
\label{fig:levels_ex0}
\end{center}
\end{figure}
%-----------------------------------------------------------
%	0+ w.f.
%-----------------------------------------------------------
\begin{figure}[ht!]
\begin{center}
\includegraphics[width=\linewidth]{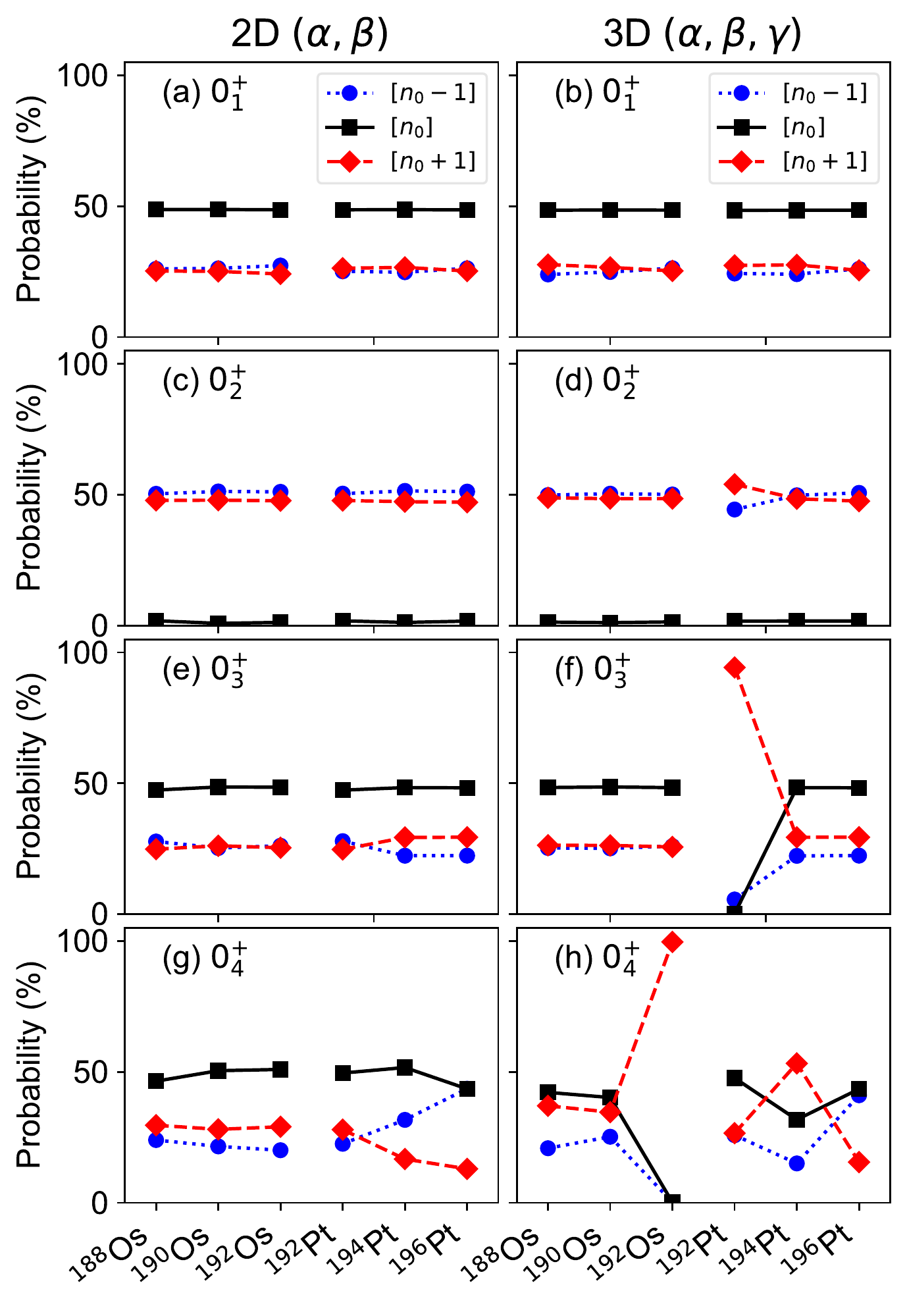}
\caption{Fractions of the $[n_{0}-1]$, $[n_{0}]$, and 
$[n_{0}+1]$ boson-space components in the wave functions of 
the four lowest $0^{+}$ excited states in 
$^{188,190,192}$Os and $^{192,194,196}$Pt. These 
values correspond to the IBM calculations based on the 
PC-PK1 EDF, and include the 2D
 (axial quadrupole $\beta$ plus pairing $\alpha$) (left column), 
and 3D (triaxial quadrupole $(\beta,\gamma)$ 
and pairing $\alpha$) (right column) degrees of freedom. 
} 
\label{fig:wf_ex0}
\end{center}
\end{figure}

%-----------------------------------------------------------
%	gamma-band energies
%-----------------------------------------------------------
\begin{figure}[ht!]
\begin{center}
\includegraphics[width=\linewidth]{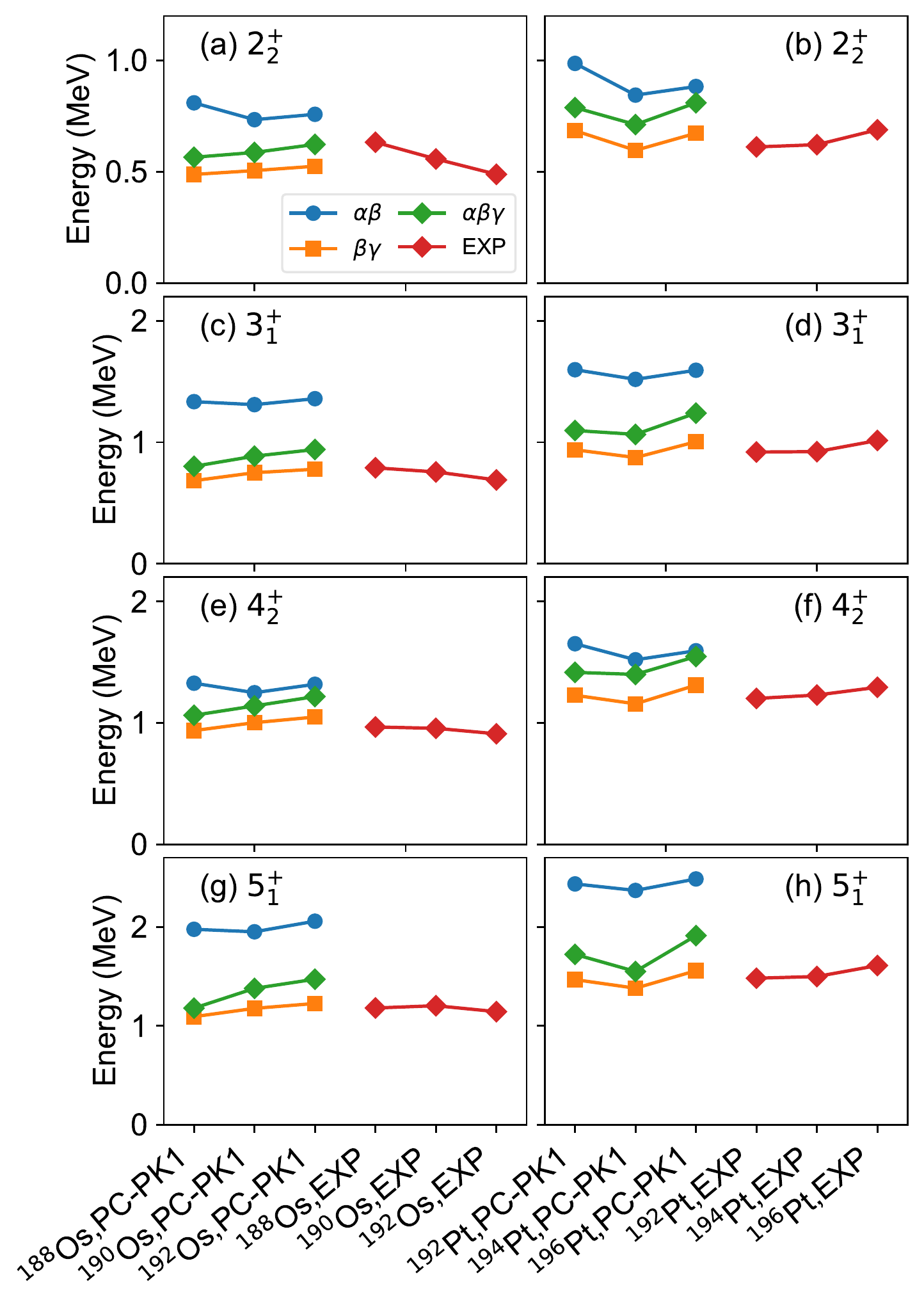}
\caption{Excitation energies of 
$\gamma$-band levels of $^{188,190,192}$Os 
and $^{192,194,196}$Pt,  obtained with the IBM
that includes the triaxial quadrupole ($\beta\gamma$), 
axial+pairing ($\alpha\beta$), and 
triaxial+pairing ($\alpha\beta\gamma$) 
degrees of freedom, in comparison with the corresponding 
experimental states. The results based on the 
functional PC-PK1 are shown.} 
\label{fig:levels_gam_pk}
\end{center}
\end{figure}
%-----------------------------------------------------------
%	S(J)
%-----------------------------------------------------------
\begin{figure}[ht!]
\begin{center}
\includegraphics[width=\linewidth]{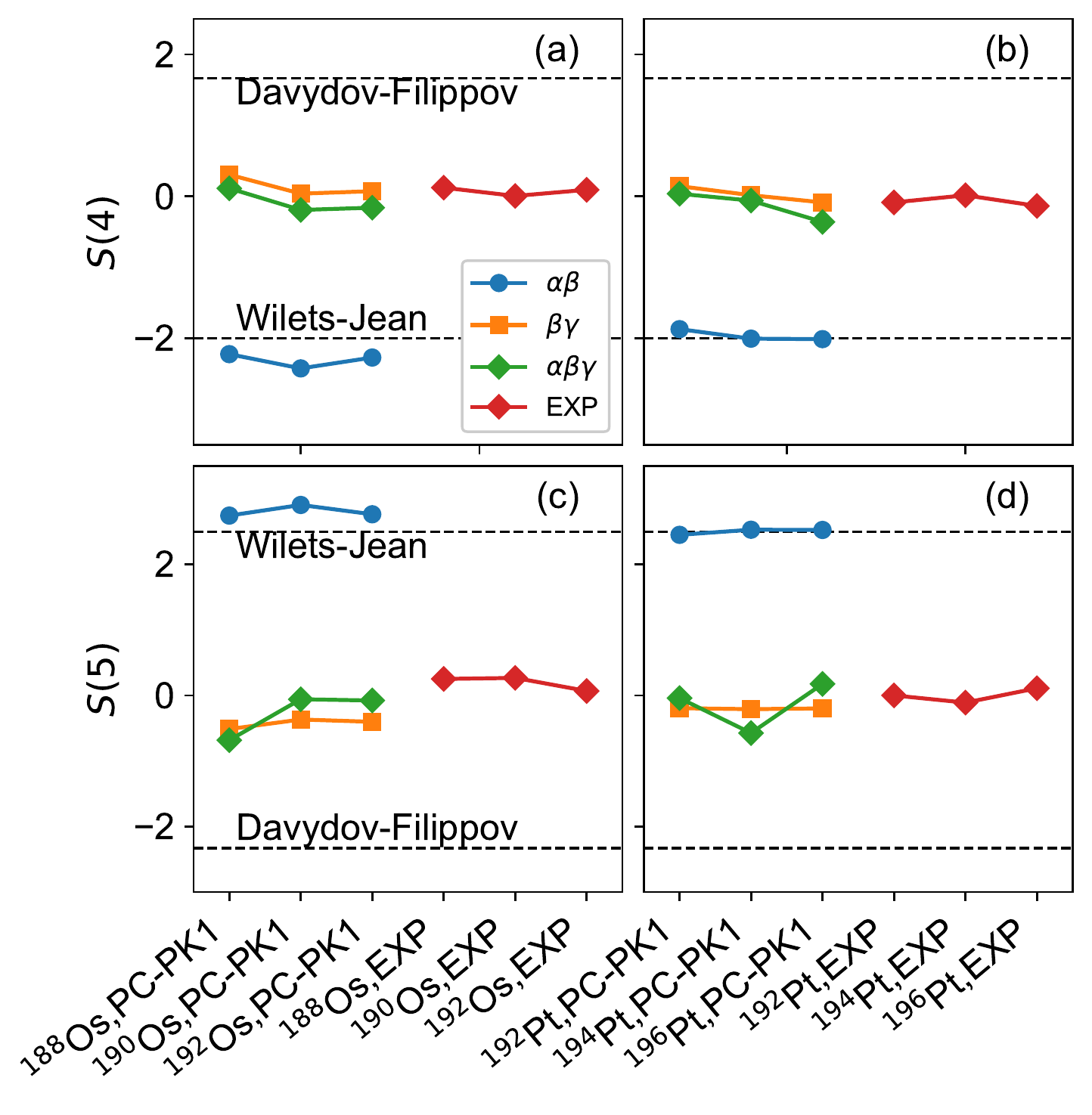}
\caption{$S(4)$ and $S(5)$ 
values for the $\gamma$-band states of Os (left) 
and Pt (right) isotopes. 
The Wilets-Jean limit: $S(4)=-2.00$ and $S(5)=2.50$, 
and the Davydov-Filippov limit: 
$S(4)=1.67$ and $S(5)=-2.33$ are indicated 
by dashed horizontal lines. The functional 
PC-PK1 is used in the IBM calculation.}
\label{fig:sj}
\end{center}
\end{figure}

%-----------------------------------------------------------
%	Comparison of PC-PK1 and DD-PC1 - Os188, Pt196
%-----------------------------------------------------------
\begin{figure}[ht!]
\begin{center}
\includegraphics[width=\linewidth]{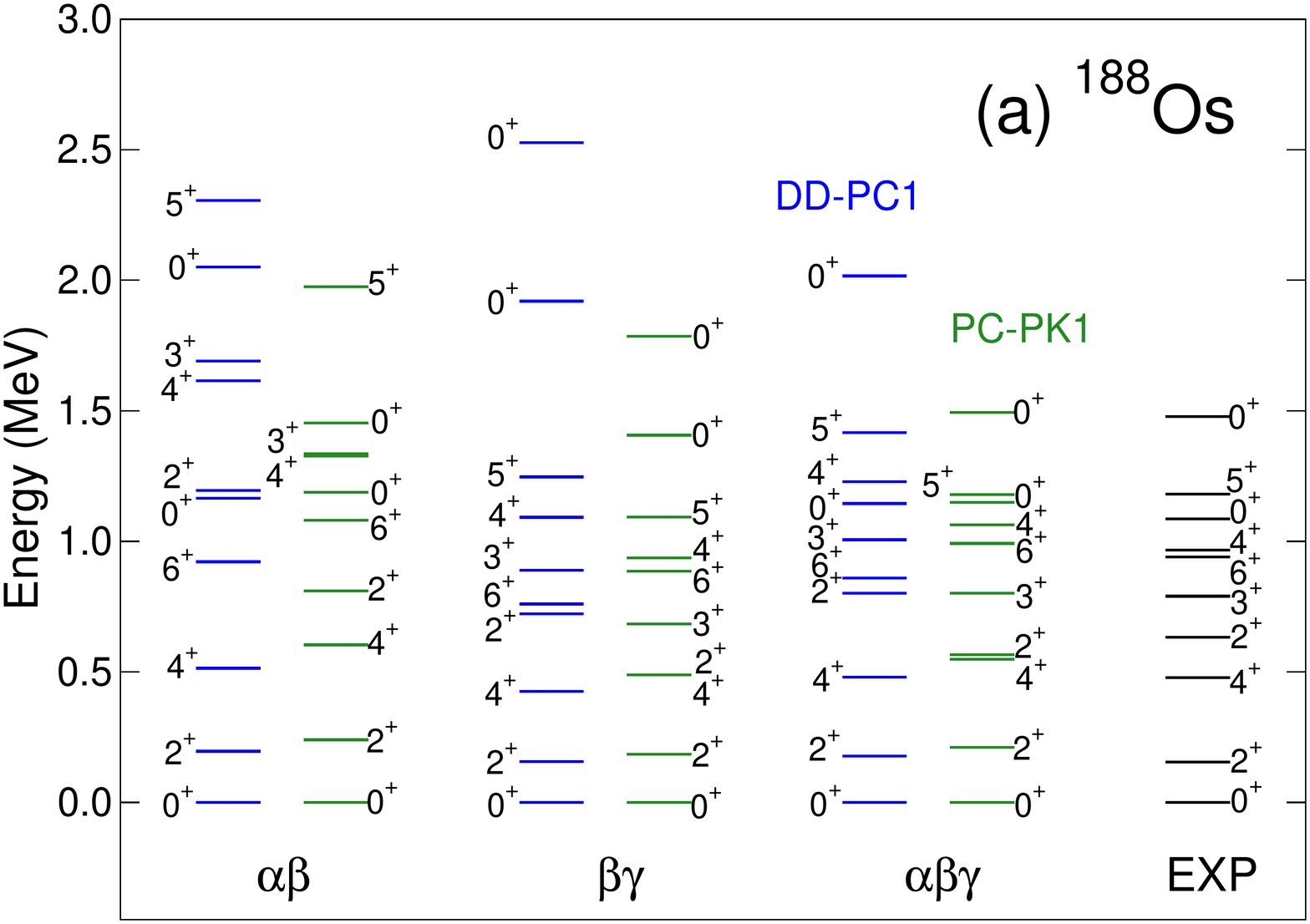}\\
\includegraphics[width=\linewidth]{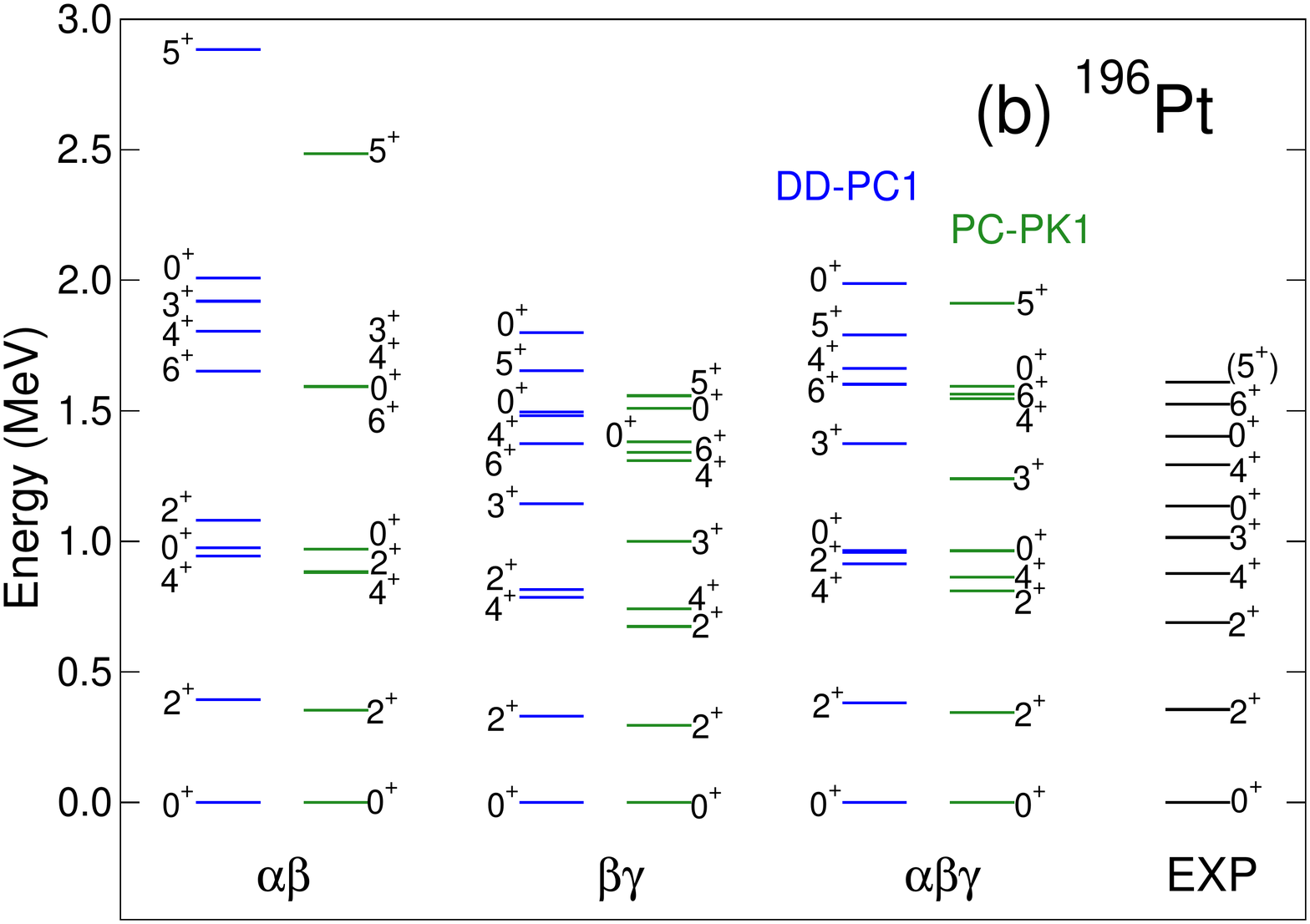}
\caption{\label{fig:comp}
The low-energy excitation spectra 
of (a) $^{188}$Os and (b) $^{196}$Pt, 
obtained from IBM calculations based on the 
DD-PC1 and PC-PK1  functionals. 
For each nucleus, results of calculations that include 
the axial+pairing ($\alpha\beta$), 
triaxial quadrupole ($\beta\gamma$), 
and triaxial+pairing ($\alpha\beta\gamma$) 
deformation degrees of freedom, are compared with experiment.} 
\end{center}
\end{figure}

\section{Spectroscopic properties\label{sec:spec}}
In the remainder of this paper, we will present   
selected spectroscopic results. Section \ref{sec:band} 
contains a discussion of low-energy band structure 
of Os and Pt isotopes, obtained 
from the 3D-IBM calculations that take 
into account both the pairing and triaxial 
degrees of freedom. 
In Secs.~\ref{sec:pairing} and \ref{sec:tri}, 
we specifically analyze the effect of including 
dynamical pairing and triaxiality on the excited 
$0^{+}$ states, and the $\gamma$-vibrational 
band, respectively. 
Section \ref{sec:comp} compares 
the excitation spectra calculated with the 
PC-PK1 and DD-PC1 energy density functionals. 
Electric transition properties 
are discussed in Sec.~\ref{sec:trans}. 

\subsection{Low-energy excitation spectra\label{sec:band}}
Figures~\ref{fig:band_all_pk} and 
\ref{fig:band_all_dd} 
display the low-energy excitation level schemes 
of the $^{188,190,192}$Os and $^{192,194,196}$Pt isotopes, 
calculated with the 3D-IBM based on the 
PC-PK1 and DD-PC1 EDFs, respectively.  
In general, the theoretical excitation spectra 
are in a good agreement with their experimental 
counterparts \cite{data}. 
Of particular interest here are the excitation energies 
of the second $0^{+}$ states, which for 
most of the considered nuclei are predicted 
very close to the experimental values. 
In the Os and Pt nuclei analyzed in the present study, 
as also shown below, the 
$0^{+}_{2}$ state exhibits  
a pairing vibrational structure, 
namely, the principal contributions to this state are from 
the $[n_{0}\pm1]$ boson subspaces. 
The model also 
predicts the third $0^{+}$ levels in agreement 
with the data, in particular for the calculation based on 
the PC-PK1 EDF. 
In many cases one notices that the energy gap 
between the calculated $0^{+}_{3}$ and $0^{+}_{4}$ 
levels is much larger than the one observed in experiment, 
and this strong repulsion between the two $0^{+}$ states 
indicates a high degree of mixing of their wave functions. 

The $\gamma$-vibrational band, 
built on the second $2^{+}$ 
state, is overall reproduced in agreement with data, 
even though the 
band-head $2^{+}_{2}$ energy is slightly  
overestimated. 
Consistent with the experimental sequence, 
in most cases the levels of the $\gamma$ band 
are almost equidistant. 
As discussed in more detail 
in Sec.~\ref{sec:tri}, 
this energy-level systematics 
is characteristic for structures that lie in between 
those predicted by the rigid-triaxial 
and $\gamma$-unstable rotor geometric models.

By comparing the results shown 
in Figs.~\ref{fig:band_all_pk} 
and \ref{fig:band_all_dd}, 
one notices that the rotational features of the 
IBM spectra resulting from the DD-PC1 EDF 
are more pronounced than for those obtained with PC-PK1.
For the functional DD-PC1, the $\gamma$ band is predicted 
to be higher with respect to the ground-state band, 
while the $0^{+}_{2}$ band for most of 
the nuclei is close to or lower than the experimental levels. 
In the IBM framework, the band-head energies and the moments of inertia 
of the $\gamma$-vibrational band and the band 
built on the $0^{+}_{2}$ state, 
are to a large extent determined by the 
magnitude of the quadrupole-quadrupole 
interaction strength $\kappa$ 
(see, Eq.~(\ref{eq:h2b})). 
As noted from Table~\ref{tab:para}, 
the value of $\kappa$  
that is used in the case of the DD-PC1 EDF 
is more than 20 \% larger than for the PC-PK1 case. 
The different IBM parameters, 
in turn, reflects the differences in 
the SCMF-PESs calculated with the two EDFs. 

The energy spectra for the Os and Pt nuclei shown 
in Figs.~\ref{fig:band_all_pk} 
and \ref{fig:band_all_dd}, are also in better agreement with 
the data when compared to previous 
IBM calculations in the Os-Pt region, 
based either on the Gogny EDFs 
\cite{nomura2011pt,nomura2011wos}, or 
the DD-PC1 EDF \cite{nomura2011coll}. 
In those studies only the two-body IBM Hamiltonian 
was considered, which resulted in the $\gamma$ band 
that exhibits a staggering pattern 
$2^{+}_{\gamma}$, $(3^{+}_{\gamma},4^{+}_{\gamma})$,
$(5^{+}_{\gamma},6^{+}_{\gamma})$, $\ldots$, 
characteristic of the $\gamma$-unstable O(6) limit.

\subsection{Effect of dynamical pairing\label{sec:pairing}}
Figure~\ref{fig:levels_ex0} displays the 
excitation energies for the second, third, 
and fourth $0^{+}$ states, obtained from 
IBM calculations that take into account the 
2D axial+pairing ($\alpha\beta$), 
2D triaxial ($\beta\gamma$), and 
3D triaxial+pairing ($\alpha\beta\gamma$) 
degrees of freedom, respectively. 
For both Os and Pt isotopes, 
the $0^{+}_{2}$ energy levels are lowered 
by a factor of $1.3-1.5$ with the 
inclusion of dynamical pairing. 
The pairing degree of freedom is also relevant 
for the description of the $0^{+}_{3}$ states 
in Os isotopes, 
while for the Pt isotopes 
it is less significant. 
Particularly with the PC-PK1 functional, 
the inclusion of dynamical pairing does not necessarily 
improve the description of the 
$0^{+}_{4}$ excitation energies. 
In the calculation based on the DD-PC1 EDF, 
dynamical pairing also 
significantly reduces the $0^{+}_{4}$ excitation 
energies. In general, it appears that dynamical pairing effects 
are more pronounced for the case in which the functional 
DD-PC1 is used as a basis of IBM calculations. 

To analyze the structure  
of $0^{+}$ states, we plot in Fig.~\ref{fig:wf_ex0} 
the percentage of the normal $[n_{0}]$ (half the number of valence nucleons), 
and pair-vibrational $[n_{0}\pm1]$ components 
in the IBM wave functions of the four 
lowest $0^{+}$ states. 
The 2D-$\alpha\beta$ and 3D-$\alpha\beta\gamma$ IBM results are 
compared in the left and right columns, respectively. 
For all the Os and Pt nuclei, 
approximately 50 \% of the wave function of the 
$0^{+}_{1}$ ground state belongs to the 
$[n_{0}]$ boson space, while the other half is equally 
shared by the $[n_{0}\pm1]$ components. 
For the $0^{+}_{2}$ state, 
as already noted above, the pair vibrational 
configurations $[n_{0}\pm1]$ dominate the 
corresponding wave function, while the contribution 
from the $[n_{0}]$ space is negligibly small. 
This conclusion is robust, in the sense that 
there is no notable difference between 
the IBM calculations with (3D-$\alpha\beta\gamma$) 
and without (2D-$\alpha\beta$) the triaxiality. 
The same conclusion 
was drawn for the $^{128,130}$Xe nuclei 
in \cite{nomura2021pv}, 
and for the axially-deformed rare-earth nuclei 
in \cite{nomura2020pv}. 
In Figs.~\ref{fig:wf_ex0}(e) to \ref{fig:wf_ex0}(h), 
the third and fourth $0^{+}$ states exhibit  
structures similar to that of 
the ground state. Some exceptions occur in the 
3D calculation 
(Figs.~\ref{fig:wf_ex0}(f) and \ref{fig:wf_ex0}(h)), 
for instance, the $0^{+}_{3}$ state of $^{192}$Pt 
and the $0^{+}_{4}$ state of $^{192}$Os. 
The irregular behavior reflects
the complexity of the 
3D calculation that involves both 
triaxial and pairing deformations. 
A similar conclusion applies to  
the DD-PC1 results.

\subsection{Effect of triaxiality\label{sec:tri}}
In Fig.~\ref{fig:levels_gam_pk}, the excitation energies 
of members of the $\gamma$-vibrational 
band of the Os and Pt nuclei are 
depicted. The theoretical values correspond 
to the 2D-$\alpha\beta$,
2D-$\beta\gamma$, and 
3D-$\alpha\beta\gamma$ IBM 
calculations, respectively, and are shown in comparison to the 
corresponding experimental levels. 
As one would expect, the pronounced effect of triaxiality 
is to lower the excitation energies of the 
$\gamma$-band states, especially the odd-spin ones. 
The energy levels of the $\gamma$ band 
obtained in the 2D-$\beta\gamma$ and 
3D-$\alpha\beta\gamma$ calculations are considerably lower 
than those resulting from the calculation that includes 
only axial $\beta$ and pairing $\alpha$ deformations. 
However, it appears that the effect of including dynamical pairing 
is to slightly raise the $\gamma$-band levels, worsening the 
agreement with available data. 
The $\gamma$-band energies computed with the 
3D-$\alpha\beta\gamma$ IBM are systematically 
higher than those obtained with the 2D-$\beta\gamma$ 
calculation. 

Two limiting geometrical pictures of non-axial nuclei 
are provided by (i) the rigid-triaxial-rotor model 
of Davydov and 
Filippov \cite{Davydov58} that corresponds to a collective 
potential with a stable minimum at a particular value 
of $\gamma$, and (ii) the $\gamma$-unstable-rotor 
model of Wilets and Jean \cite{gsoft}  
that describes a collective 
potential that is independent of $\gamma$. 
To distinguish between the energy-level structure 
of $\gamma$-vibrational bands in the two geometrical limits, 
we consider the quantity $S(I)$, 
defined in terms of the excitation 
energies of the members of a $\gamma$ band:
\begin{align}
 S(I_{\gamma}) =
 \frac{(E(I_{\gamma})-E(I_{\gamma}-1))-(E(I_{\gamma}-1)-E(I_{\gamma}-2))}
{E(2^{+}_{1})}
\end{align}
In the ideal $\gamma$-unstable-rotor case, 
its values for $I_{\gamma}=4$ and $I_{\gamma}=5$  
are $S(4)=-2.00$ and $S(5)=2.50$, respectively. 
%(in the following, the subscript $\gamma$ with 
%$I_{\gamma}$ is omitted). 
These values reflect an approximate grouping pattern 
$2^{+}_{\gamma}$, $(3^{+}_{\gamma},4^{+}_{\gamma})$,
$(5^{+}_{\gamma},6^{+}_{\gamma})$, $\ldots$ etc. 
In the rigid-triaxial-rotor limit, 
the $S(I)$ values are: $S(4)=1.67$ and $S(5)=-2.33$, 
corresponding to the staggering 
$(2^{+}_{\gamma},3^{+}_{\gamma})$,
$(4^{+}_{\gamma},5^{+}_{\gamma})$, $\ldots$ etc.

Figure~\ref{fig:sj} depicts the values of 
$S(4)$ and $S(5)$ for 
the Os and Pt nuclei considered in the present study. 
The 2D-$\alpha\beta$ IBM calculation
in all cases predicts the $S(4)$ and $S(5)$ values 
close to the $\gamma$-unstable-rotor limit 
or the O(6) limit of the IBM \cite{IBM}. 
This is  because any IBM-1 Hamiltonian that includes only 
two-body boson terms does not give rise to a triaxial minimum, 
and the resulting $\gamma$-band is always that of 
the $\gamma$-unstable-rotor. 
When triaxiality is 
taken into account in the 2D-$\beta\gamma$ and 
3D-$\alpha\beta\gamma$ calculations, and including  
the three-body boson term (\ref{eq:h3b}), 
both $S(4)\approx S(5)\approx 0$, a value that is
almost halfway between the two geometrical 
limits. 
We note that the calculated $S(4)$ and $S(5)$ values 
in the 2D-$\beta\gamma$ and 3D-$\alpha\beta\gamma$ 
calculations are in very good agreement with the 
corresponding experimental values. 
The same conclusion was reached in our previous 
study of $^{128,130}$Xe \cite{nomura2021pv}. 
The results for the $\gamma$-band obtained with the DD-PC1 EDF are 
quantitatively similar to those shown 
in Figs.~\ref{fig:levels_gam_pk} and \ref{fig:sj}.

\subsection{The DD-PC1 and PC-PK1 excitation spectra \label{sec:comp}}
Figure~\ref{fig:comp} 
compares the low-energy excitation spectra,  
calculated using the DD-PC1 and PC-PK1 EDFs to 
determine the parameters of the IBM Hamiltonian. 
As examples, we consider the nuclei $^{188}$Os and $^{196}$Pt. 
A common feature of the two EDFs is 
that, by the inclusion of dynamical pairing, the 
excited $0^{+}$ states are considerably lowered. 
Also, when the triaxial degree of freedom 
is taken into account (that is, the three-body 
boson interaction is included), 
states belonging to the $\gamma$-band, 
especially the odd-spin members, are predicted at 
significantly lower excitation energies. 
On a closer inspection, it appears that the 3D-IBM calculation 
based on the PC-PK1 EDF produces results in better 
quantitative agreement with experiment. 
As already noted in Sec.~\ref{sec:band}, 
the IBM calculation based on the 
DD-PC1 functional generally lead to
energy spectra that are stretched 
compared to the PC-PK1 model calculation.

\subsection{Transition rates\label{sec:trans}}

\subsubsection{The $E2$ and $E0$ operators}
The electric quadrupole ($E2$) and monopole ($E0$) 
transition rates will also be influenced by the simultaneous inclusion of 
the quadrupole triaxial and pairing degrees of freedom. An accurate 
description of these transition rates thus provides a stringent test of the model. 
The general (one-body) 
$E2$ and $E0$ operators are defined by the following relations: 
\begin{align}
\label{eq:e2}
& \hat T(E2)  = e_\mathrm{B}\hat{Q} \\
\label{eq:e0}
&\hat T(E0)  = \xi\hat n_d + \zeta\hat n
\end{align}
where $e_\mathrm{B}$ is the $E2$ boson effective charge, 
$\hat{Q}$ is the same quadrupole operator that appears 
in the boson Hamiltonian (\ref{eq:h2b}), 
and $\xi$ and $\zeta$ in the $E0$ operators 
denote parameters. 
The $B(E2)$ and $\rho^2(E0)$ transition rates are  
calculated using the expressions:
\begin{align}
&
B(E2; I_i\to I_f)
=\frac{1}{2I_i+1} | \braket{I_f \| \hat T(E2) \| I_i} |^2 \\
&
\rho^2(E0; I_i\to I_f)
=\frac{Z^2}{e^2r_0^4A^{4/3}} \frac{1}{2I_i + 1}| \braket{I_f \| \hat T(E0) \| I_i} |^2 \;.
\end{align}
The boson charge $e_{\mathrm{B}}=0.145$ $e$b is adjusted and kept constant  
for all the Os and Pt isotopes, and for 
both the PC-PK1 and DD-PC1 EDFs. This choice for the boson charge  ensures 
that the experimental $B(E2;2^{+}_{1}\to 0^{+}_{1})$ 
values are reasonably 
reproduced. In the lighter mass region, $e_{\mathrm{B}}=0.12$ $e$b 
is used for $^{128,130}$Xe. 
The $E0$ parameters $\xi=0.112$ (0.093) 
and $\zeta=-0.09$ ($-0.062$) fm$^2$ are used for Xe, Os, 
and Pt nuclei, in calculations with the PC-PK1 (DD-PC1) EDF. 
These values are determined to 
reproduce the experimental 
$\rho^{2}(E0;0^{+}_{2}\to 0^{+}_{1})$ 
values of $^{188}$Os and $^{194}$Pt 
in the 3D-IBM calculation. 

It must be noted that, since the boson 
Hamiltonian consists 
of up to three-body boson terms,
in the calculations that include triaxial 
degrees of freedom, i.e., the $(\beta,\gamma)$- 
and $(\alpha,\beta,\gamma)$-IBM, 
the $E2$ operator should also 
contain higher-order terms: 
\begin{align}
\label{eq:e2-tb}
\hat{T}(E2)=&e_{sd}(s^\+\tilde d + d^\+s) + e_{dd} 
(d^\+\tilde d)^{(2)} \nonumber \\
& + e_{sddd}\sum_{L=0,2,4}\{[(s^\+\tilde
d)(d^\+\tilde d)^{(L)}]^{(2)} + (h.c.)\} \nonumber \\
& + e_{dddd}\sum_{L,L'=0,2,4}[(d^\+d^\+)^{(L)}(\tilde d\tilde
d)^{(L')}]^{(2)}. 
\end{align}
with additional boson effective charges $e_{sd}$, $e_{dd}$,
$e_{sddd}$, and $e_{dddd}$. 
The $B(E2)$ transition rates calculated by using the 
two-body $E2$ operator 
for the $^{128,130}$Xe 
can be found in Fig.~4 of Ref.~\cite{nomura2021pv}. 
The $E2$ charges for the operator ($\ref{eq:e2-tb}$) 
chosen for $^{128}$Xe ($^{130}$Xe) 
in \cite{nomura2021pv} were 
$e_{sd}=0.078$ (0.09), $e_{dd}=0.034$ (0.086),
$e_{sddd}=0.016$ (0.015), and $e_{dddd}=0.009$ (0) $e$b. 
To keep the calculation and discussion as simple as possible, 
in this work we use the standard 
$E2$ operator in (\ref{eq:e2}) 
for all considered nuclei. 

In addition, 
%as is often made in the 
%configuration-mixing IBM calculations 
%\cite{heyde2011,nomura2016sc}, 
the $E2$ effective boson charge $e_{\mathrm{B}}$ 
and the parameters $\xi$ and $\zeta$ in the $E0$ operator 
could, in principle, differ in the three boson subspaces 
$[n_{0}]$ and $[n_{0}\pm 1]$. Again, for simplicity,
we use the same values of these parameters 
for the three configuration spaces, just as in the case of 
the Hamiltonian parameters. 

%-----------------------------------------------------------------------
% B(E2) - Xe, PC-PK1
%-----------------------------------------------------------------------
\begin{table}[!htb]
\begin{center}
\caption{\label{tab:e2-xe}
Comparison of experimental 
\cite{data} and theoretical $B(E2;I_{i}\to I_{f})$ values 
(in Weisskopf units) for $^{128,130}$Xe. 
The theoretical values are obtained from 
IBM calculations that include triaxial quadrupole 
(denoted by $\beta\gamma$), 
axial plus dynamical pairing ($\alpha\beta$), 
and triaxial plus dynamical pairing ($\alpha\beta\gamma$) 
degrees of freedom. 
The numbers in parentheses for the 2D-$\beta\gamma$ 
and 3D-$\alpha\beta\gamma$ calculations 
denote values that are obtained using the 
two-body $E2$ operator in Eq.~(\ref{eq:e2-tb}). 
The IBM calculations are based on the 
PC-PK1 energy density functional.}
 \begin{ruledtabular}
 \begin{tabular}{lcccccc}
  & $I_{i}$ & $I_{f}$ & EXP & $\beta\gamma$ & $\alpha\beta$ & $\alpha\beta\gamma$ \\ 
\hline
$^{128}$Xe
& $2^+_1$ & $0^+_1$ & 42.6$\pm6.4$ & 45 (50) & 44 & 46 (49) \\
& $4^+_1$ & $2^+_1$ & 63.5$\pm5.2$ & 59 (85) & 60 & 61 (82) \\
& $6^+_1$ & $4^+_1$ & 106$\pm13$ & 60 (95) & 62 & 65 (96) \\
& $2^+_2$ & $2^+_1$ & 50.1$\pm9.7$ & 54 (48) & 25 & 57 (52) \\
&         & $0^+_1$ & 0.65$\pm0.08$ & 0.29 (11) & 2.6 & 0.22 (9.8)\\
& $3^+_1$ & $4^+_1$ & 31.8$\pm5.9$ & 20 (19) & 11 & 23 (23) \\
&         & $2^+_2$ & 91$\pm16$ & 49 (63) & 12 & 57 (75) \\
&         & $2^+_1$ & 1.45$\pm0.26$ & 0.39 (8.1) & 37 & 0.33 (9.6) \\
& $4^+_2$ & $4^+_1$ & 30.2$\pm3.2$ & 23 (31) & 6.5 & 21 (30) \\
&         & $2^+_2$ & 29.6$\pm2.9$ & 28 (23) & 21 & 27 (28) \\
&         & $2^+_1$ & 0.52$\pm0.06$ & 0.0056 (54) & 0.042 & 0.020 (47) \\
& $0^+_2$ & $2^+_2$ & 52.8$\pm0.7.6$ & 40 (40) & 30 & 0.19 (19) \\
&         & $2^+_1$ & 3.69$\pm0.58$ & 0.37 (115) & 4.0 & 3.8 (16) \\
& $0^+_3$ & $2^+_2$ & 22.2$\pm4.6$ & 0.37 (73) & 33 & 35 (38) \\
&         & $2^+_1$ & 10.4$\pm2.3$ & 0.28 (12) & 0.62 & 0.30 (101) \\
\hline
$^{130}$Xe
& $2^+_1$ & $0^+_1$ & 33.2$\pm 2.6$ & 33 (42) & 32 & 33 (35) \\
& $4^+_1$ & $2^+_1$ & 46.4$\pm 4.6$ & 42 (70) & 43 & 42 (60) \\
& $6^+_1$ & $4^+_1$ & 69$\pm 9$ & 41 (78) & 43 & 42 (73) \\
& $2^+_2$ & $2^+_1$ & 40$^{+10}_{-7}$ & 35 (39) & 16 & 38 (42) \\
&         & $0^+_1$ & 0.27$^{+0.07}_{-0.05}$ & 0.34 (4.1) & 1.8 & 0.18 (3.4) \\
& $3^+_1$ & $4^+_1$ & $<270$ & 13 (16) & 7.5 & 14 (19) \\
&         & $2^+_2$ & 57$^{+59}_{-42}$ & 33 (56) & 22 & 37 (61) \\
&         & $2^+_1$ & 1.0$^{+1.0}_{-0.7}$ & 0.43 (2.4) & 3.4 & 0.26 (2.0) \\
& $4^+_2$ & $4^+_1$ & 42$^{+48}_{-35}$ & 14 (25) & 5.1 & 14 (26) \\
&         & $2^+_2$ & 69$^{+65}_{-57}$ & 17 (21) & 17 & 18 (15) \\
&         & $2^+_1$ & 2.1$^{+2.0}_{-1.7}$ & 0.016 (21) & 0.024 & 0.029 (22) \\
& $0^+_2$ & $2^+_2$ & 120$^{+110}_{-70}$ & 23 (48) & 36 & 0.39 (19) \\
&         & $2^+_1$ & 18$^{+17}_{-11}$ & 0.65 (78) & 4.5 & 2.7 (19) \\
& $0^+_3$ & $2^+_2$ & 55$^{+37}_{-33}$ & 2.5 (6.1) & 5.6 & 0.45 (1.3) \\
&         & $2^+_1$ & & 0.17 (6.1) & 0.024 & 1.5 (7.2) \\
 \end{tabular}
 \end{ruledtabular}
\end{center} 
\end{table}
%-----------------------------------------------------------------------
% B(E2) - Os, PC-PK1
%-----------------------------------------------------------------------
\begin{table}[!htb]
\begin{center}
\caption{\label{tab:e2-os}
Same as in the caption to Table~\ref{tab:e2-xe} but for the Os nuclei. 
Only results calculated with the one-body $E2$ operator are shown.}
 \begin{ruledtabular}
 \begin{tabular}{lcccccc}
  & $I_{i}$ & $I_{f}$ & EXP & $\beta\gamma$ & $\alpha\beta$ & $\alpha\beta\gamma$ \\ 
\hline
$^{188}$Os 
& $2^+_1$ & $0^+_1$ & 77.5$\pm1.0$ & 88 & 89 & 89 \\
& $4^+_1$ & $2^+_1$ & 133$\pm8$ & 123 & 125 & 125 \\
& $6^+_1$ & $4^+_1$ & 138$\pm8$ & 139 & 139 & 142 \\
& $2^+_2$ & $2^+_1$ & 16.2$\pm1.8$ & 80 & 81 & 85 \\
&         & $0^+_1$ & 5.0$\pm0.6$ & 3.1 & 3.0 & 2.9 \\
& $3^+_1$ & $4^+_1$ & & 52 & 32 & 51 \\
&         & $2^+_2$ & & 129 & 97 & 130 \\
&         & $2^+_1$ & & 5.0 & 4.6 & 4.7 \\
& $4^+_2$ & $6^+_1$ & 15$\pm6$ & 2.0 & 1.0 & 1.1 \\
&         & $4^+_1$ & 19$\pm3$ & 35 & 49 & 22 \\
&         & $3^+_1$ & & 27 & 11 & 14 \\
&         & $2^+_2$ & 47$\pm8$ & 45 & 72 & 29 \\
&         & $2^+_1$ & 1.31$\pm0.19$ & 0.4 & 0.03 & 0.14 \\
& $0^+_2$ & $2^+_2$ & 4.8$\pm0.3$ & 36 & 5.4 & $7.0\times10^{-4}$ \\
&         & $2^+_1$ & 0.96$\pm0.05$ & 1.5 & 2.9 & 2.6 \\
& $0^+_3$ & $2^+_2$ &  & $2.3\times10^{-5}$ & 125 & 46 \\
&         & $2^+_1$ &  & 0.016 & 2.2 & 1.4 \\
\hline
$^{190}$Os 
& $2^+_1$ & $0^+_1$ & 72.9$\pm1.6$ & 75 & 75 & 75 \\
& $4^+_1$ & $2^+_1$ & 99$^{+5}_{-3}$ & 102 & 103 & 103 \\
& $6^+_1$ & $4^+_1$ & 113$\pm7$ & 114 & 113 & 115 \\
&         & $4^+_2$ & 5.6$^{+4.5}_{-3.6}$ & 0.051 & 0.023 & 0.041 \\
& $2^+_2$ & $2^+_1$ & 32.6$\pm3.4$ & 99 & 101 & 101 \\
&         & $0^+_1$ & 6.0$\pm0.6$ & 0.12 & 0.10 & 0.10 \\
& $3^+_1$ & $4^+_1$ & & 46 & 32 & 44 \\
&         & $2^+_2$ & & 103 & 80 & 103 \\
&         & $2^+_1$ & & 0.19 & 0.15 & 0.16 \\
& $4^+_2$ & $6^+_1$ & & 0.074 & 0.033 & 0.060 \\
&         & $4^+_1$ & 31$\pm5$ & 36 & 53 & 40 \\
&         & $3^+_1$ & 54$^{+24}_{-19}$ & 0.93 & 0.33 & 0.72 \\
&         & $2^+_2$ & 52.3$\pm4.3$ & 47 & 59 & 50 \\
&         & $2^+_1$ & 0.69$\pm0.06$ & 0.018 & 0.0002 & 0.009 \\
& $0^+_2$ & $2^+_2$ & 24$^{+10}_{-7}$ & 47 & 0.20 & 0.012 \\
&         & $2^+_1$ & 2.4$^{+0.8}_{-0.6}$ & 0.078 & 2.0 & 2.5 \\
& $0^+_3$ & $2^+_2$ &  & 2.2 & 0.016 & 0.0007 \\
&         & $2^+_1$ &  & 125 & $2.3\times10^{-5}$ & 2.6 \\
\hline
$^{192}$Os 
& $2^+_1$ & $0^+_1$ & 62.1$\pm0.7$ & 61 & 60 & 61 \\
& $4^+_1$ & $2^+_1$ & 75.6$\pm2.0$ & 81 & 82 & 82 \\
& $6^+_1$ & $4^+_1$ & 100$^{+5}_{-3}$ & 89 & 88 & 90 \\
& $2^+_2$ & $2^+_1$ & 46.0$^{+2.6}_{-1.2}$ & 79 & 81 & 81 \\
&         & $0^+_1$ & 5.62$^{+0.21}_{-0.12}$ & 0.072 & 0.055 & 0.056 \\
& $3^+_1$ & $4^+_1$ & 36 & 25 & 34 \\
&         & $2^+_2$ & 81 & 63 & 80 \\
&         & $2^+_1$ & 0.11 & 0.079 & 0.085 \\
& $4^+_2$ & $6^+_1$ & & 0.043 & 0.015 & 0.032 \\
&         & $4^+_1$ & 30.9$^{+3.6}_{-1.8}$ & 28 & 42 & 32 \\
&         & $3^+_1$ & & 0.61 & 0.16 & 0.44 \\
&         & $2^+_2$ & 45.2$^{+1.4}_{-1.8}$ & 36 & 46 & 39 \\
&         & $2^+_1$ & 0.29$\pm0.03$ & 0.013 & $2.0\times10^{-5}$ & 0.0053 \\
& $0^+_2$ & $2^+_2$ & 30.4$^{+3.0}_{-2.3}$ & 37 & 0.097 & 0.010 \\
&         & $2^+_1$ & 0.57$\pm0.12$ & 0.050 & 2.0 & 2.5 \\
& $0^+_3$ & $2^+_2$ & & 0.0017 & 88 & 45 \\
&         & $2^+_1$ & 0.24$\pm0.09$ & $3.3\times10^{-5}$ & 0.052 & 0.038 \\
 \end{tabular}
 \end{ruledtabular}
\end{center} 
\end{table}
%-----------------------------------------------------------------------
% B(E2) - Pt, PC-PK1
%-----------------------------------------------------------------------
\begin{table}[!htb]
\begin{center}
\caption{\label{tab:e2-pt}
Same as in the caption to Table~\ref{tab:e2-os} but for the Pt nuclei.}
 \begin{ruledtabular}
 \begin{tabular}{lcccccc}
  & $I_{i}$ & $I_{f}$ & EXP & $\beta\gamma$ & $\alpha\beta$ & $\alpha\beta\gamma$ \\ 
\hline
$^{192}$Pt 
& $2^+_1$ & $0^+_1$ & 57.2$\pm1.2$ & 59 & 59 & 60 \\
& $4^+_1$ & $2^+_1$ & 89$\pm5$ & 82 & 83 & 83 \\
& $6^+_1$ & $4^+_1$ & 70$\pm30$ & 89 & 90 & 91 \\
& $2^+_2$ & $2^+_1$ & 109$\pm7$ & 39 & 44 & 44 \\
&         & $0^+_1$ & 0.55$\pm0.04$ & 2.7 & 2.2 & 2.4 \\
& $3^+_1$ & $4^+_1$ & 38$\pm10$ & 25 & 18 & 25 \\
&         & $2^+_2$ & 102$\pm10$ & 80 & 61 & 79 \\
&         & $2^+_1$ & 0.68$\pm0.07$ & 4.1 & 3.5 & 3.6 \\
& $4^+_2$ & $4^+_1$ & & 21 & 27 & 24 \\
&         & $2^+_2$ & & 29 & 44 & 33 \\
&         & $2^+_1$ & & 0.097 & 0.0021 & 0.054 \\
& $0^+_2$ & $2^+_2$ & & 26 & 6.4 & 0.45 \\
&         & $2^+_1$ & & 1.8 & 3.3 & 3.0 \\
& $0^+_3$ & $2^+_2$ & & 0.47 & 74 & 0.002 \\
&         & $2^+_1$ & & 0.14 & 2.2 & 0.40 \\
\hline
$^{194}$Pt 
& $2^+_1$ & $0^+_1$ & 49.2$\pm0.8$ & 48 & 48 & 48 \\
& $4^+_1$ & $2^+_1$ & 85$\pm5$ & 63 & 64 & 65 \\
& $6^+_1$ & $4^+_1$ & 67$\pm21$ & 67 & 68 & 70 \\
& $2^+_2$ & $2^+_1$ & 89$\pm11$ & 61 & 64 & 63 \\
&         & $0^+_1$ & 0.29$\pm0.04$ & 0.072 & 0.052 & 0.056 \\
& $3^+_1$ & $4^+_1$ & & 27 & 19 & 27 \\
&         & $2^+_2$ & & 62 & 49 & 63 \\
&         & $2^+_1$ & & 0.11 & 0.077 & 0.084 \\
& $4^+_2$ & $4^+_1$ & 14 & 15 & 32 & 23 \\
&         & $2^+_2$ & 21$\pm4$ & 20 & 36 & 29 \\
&         & $2^+_1$ & 0.36$\pm0.07$ & 0.0044 & $1.1\times10^{-5}$ & 0.0036 \\
& $0^+_2$ & $2^+_2$ & 8.4$\pm1.9$ & 27 & 0.12 & 0.016 \\
&         & $2^+_1$ & 0.63$\pm0.14$ & 0.060 & 2.2 & 2.8 \\
& $0^+_3$ & $2^+_2$ & & 0.11 & 68 & 34 \\
&         & $2^+_1$ & & 0.0082 & 0.065 & 0.052 \\
& $0^+_4$ & $2^+_2$ & 14.3$\pm1.4$ & 0.015 & 0.056 & 0.0043 \\
&         & $2^+_1$ & 14.1$\pm1.2$ & 0.44 & 0.21 & 0.042 \\
\hline
$^{196}$Pt 
& $2^+_1$ & $0^+_1$ & 40.60$\pm0.20$ & 37 & 36 & 37 \\
& $4^+_1$ & $2^+_1$ & 60.0$\pm0.9$ & 48 & 48 & 48 \\
& $6^+_1$ & $4^+_1$ & 73$^{+4}_{-73}$ & 50 & 50 & 51 \\
& $2^+_2$ & $2^+_1$ &  & 47 & 48 & 48 \\
&         & $0^+_1$ & $(4\pm4)\times10^{-6}$ & 0.036 & 0.021 & 0.023 \\
& $3^+_1$ & $4^+_1$ & & 18 & 14 & 17 \\
&         & $2^+_2$ & & 43 & 35 & 43 \\
&         & $2^+_1$ & & 0.049 & 0.031 & 0.034 \\
& $4^+_2$ & $4^+_1$ & 17$\pm6$ & 18 & 23 & 20 \\
&         & $2^+_2$ & 29$^{+6}_{-29}$ & 22 & 26 & 23 \\
&         & $2^+_1$ & 0.56$^{+0.12}_{-0.17}$ & 0.0042 & $5.0\times10^{-4}$ & 0.0020 \\
& $0^+_2$ & $2^+_2$ & 18$\pm10$ & 28 & 0.078 & 0.032 \\
&         & $2^+_1$ & 2.8$\pm1.5$ & 0.044 & 2.4 & 2.8 \\
& $0^+_3$ & $2^+_2$ & $<0.41$ & 0.011 & 49 & 33 \\
&         & $2^+_1$ & $<5.0$ & 0.0073 & 0.043 & 0.040 \\
 \end{tabular}
 \end{ruledtabular}
\end{center} 
\end{table}
%-----------------------------------------------------------
%	E2 ratios
%-----------------------------------------------------------
\begin{figure}[ht!]
\begin{center}
\includegraphics[width=\linewidth]{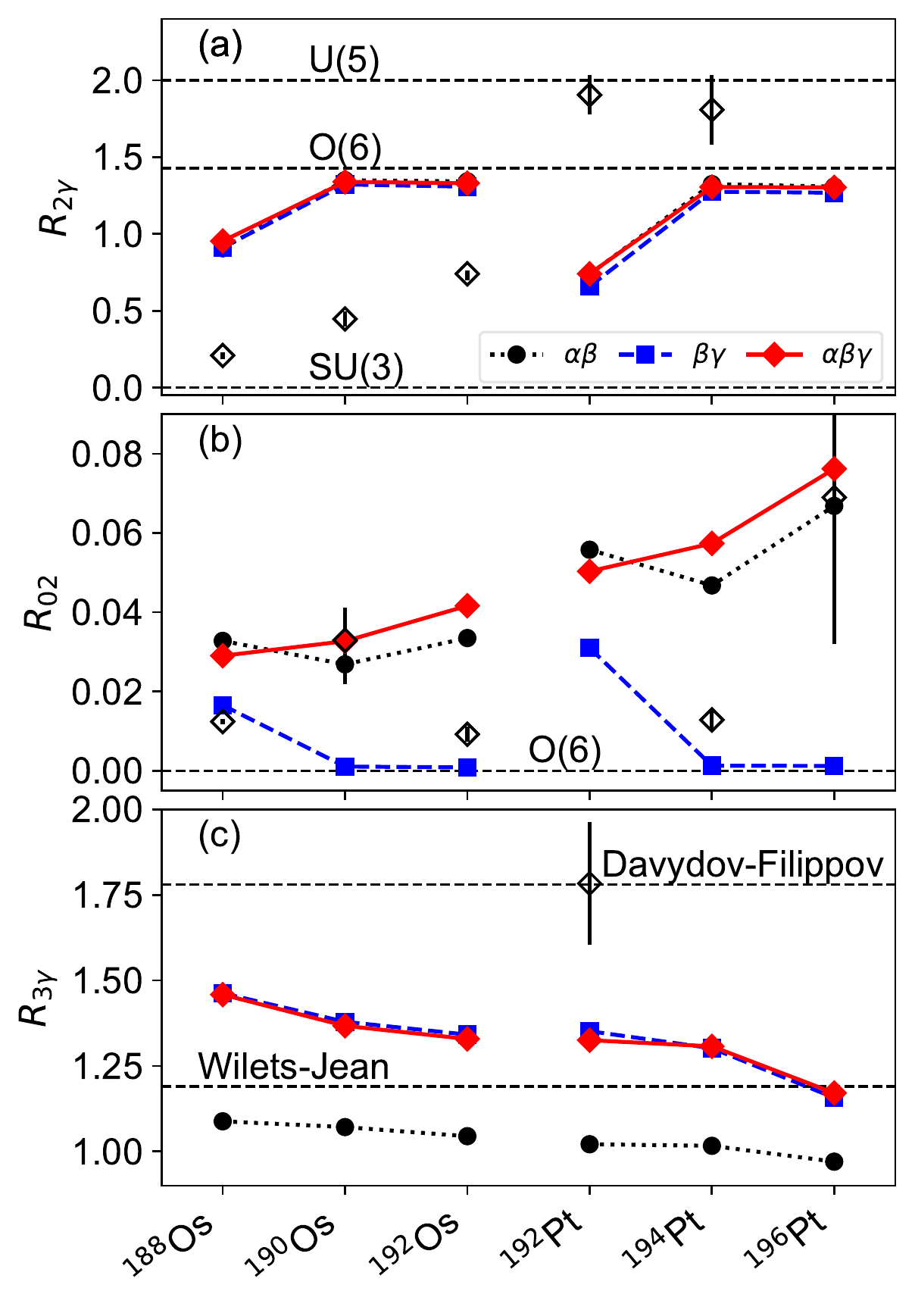}
\caption{The $B(E2)$ ratios 
(a) $R_{2\gamma}\equiv 
B(E2;2^{+}_{\gamma}\to 2^{+}_{g})/B(E2;2^{+}_{g}\to 0^{+}_{g})$, 
(b) $R_{02}\equiv 
B(E2;0^{+}_{2}\to 2^{+}_{g})/B(E2;2^{+}_{g}\to 0^{+}_{g})$, 
and 
(c) $R_{3\gamma}\equiv 
B(E2;3^{+}_{\gamma}\to 2^{+}_{g})/B(E2;2^{+}_{g}\to 0^{+}_{g})$
for  $^{188,190,182}$Os and $^{192,194,196}$Pt.
The IBM calculation is based on the PC-PK1 functional. 
The dynamical symmetry limits are indicated in panels (a) and (b). 
In panel (c), the geometrical limits of the Wilets-Jean 
$R_{3\gamma}=1.19$, and the Davydov-Filippov 
models $R_{3\gamma}=1.78$ are denoted by dashed lines. 
Experimental results are taken from \cite{data}, and are 
represented by the open symbols.} 
\label{fig:e2}
\end{center}
\end{figure}

\subsubsection{$E2$ transition properties\label{sec:e2}}
For the Xe, Os, and Pt nuclei considered in this study, 
a wealth of experimental information on 
 $E2$ transition rates is available. 
Tables~\ref{tab:e2-xe}, \ref{tab:e2-os}, and 
\ref{tab:e2-pt} compare the experimental 
\cite{data} and theoretical 
$B(E2)$ reduced transition probabilities (in Weisskopf units)
for the Xe, Os, and Pt isotopes, respectively. 
In the tables the theoretical values correspond to 
IBM calculations that include triaxial quadrupole 
($\beta\gamma$), 
axial plus dynamical pairing ($\alpha\beta$), 
and triaxial plus dynamical pairing ($\alpha\beta\gamma$) 
degrees of freedom. 

In Table~\ref{tab:e2-xe}, it is interesting to note  
that the inclusion of the pairing degree of freedom has the effect 
of slightly increasing the $B(E2;0^{+}_{2}\to 2^{+}_{1})$ 
values for both $^{128}$Xe and $^{130}$Xe. 
This is seen from the comparison of the 
2D-$\beta\gamma$ and the 
3D-$\alpha\beta\gamma$ results. 
This increase in the $E2$ transition rates is
a consequence of configuration mixing between 
the three boson subspaces, resulting in a larger 
overlap between initial and final state wave functions.  
By comparing the 2D-$\alpha\beta$ results with 
those obtained from the 2D-$\beta\gamma$ or 
3D-$\alpha\beta\gamma$ IBM calculations, 
one notices that the inclusion of the triaxial degree of freedom 
has a marked effect on the $E2$ rates that involve 
members of the $\gamma$ band. The most 
prominent example is the increase of the 
$B(E2;3^{+}_{1}\to 2^{+}_{2})$ values. 
Note that, 
in the 3D-$\alpha\beta\gamma$ IBM results 
for $^{130}$Xe, the low-spin members of 
the $\gamma$ band are the $2^{+}_{2}$, 
$3^{+}_{1}$, $4^{+}_{3}$, and $5^{+}_{1}$ levels. 
Thus in Table~\ref{tab:e2-xe}, 
the experimental $B(E2;4^{+}_{2}\to I_{f})$ 
are actually compared 
with the theoretical $B(E2;4^{+}_{3}\to I_{f})$ 
values. 
For the ground-state band, 
neither triaxiality nor dynamical pairing 
degree of freedom 
has any effect on the in-band $E2$ transition rates.

For $^{128,130}$Xe, 
the $B(E2)$ values for the 2D-$\beta\gamma$ and 
3D-$\alpha\beta\gamma$ IBM calculations that 
employ the two-body $E2$ operator (\ref{eq:e2-tb}) 
are also included (values in parentheses in 
Table~\ref{tab:e2-xe}). These values are from 
Ref.~\cite{nomura2021pv}. 
With the inclusion of the higher-order terms and hence 
additional adjustable parameters, the $B(E2)$ values 
calculated with the two-body $E2$ operator (\ref{eq:e2-tb}), 
in some cases agree better with the 
data, compared to those obtained with only the one-body $E2$ operator 
(\ref{eq:e2}), e.g., the 3D-IBM results for the 
$E2$ transitions of the $0^{+}_{2}$ states in 
both $^{128}$Xe and $^{130}$Xe. 
Nevertheless, it appears that  
the two-body $E2$ operator does not 
improve dramatically the overall description of 
$B(E2)$ transition probabilities. One should also 
note that the experimental results for these transitions 
have very large error bars. 

Similar observations can be made for the 
Os (Table~\ref{tab:e2-os}) and Pt 
(Table~\ref{tab:e2-pt}) nuclei. 
That is, the inclusion of dynamical pairing has a major 
effect on those $B(E2)$ values 
that are related to the $0^{+}_{2}$ 
state, while triaxiality leads to a better 
description of the $E2$ rates for $\gamma$-band states. 
For all Os and Pt nuclei except $^{194}$Pt, 
in the 3D-IBM calculations the second $4^{+}$ state 
corresponds to the $I=4^{+}$ member of the 
$\gamma$-vibrational band. 
Therefore, for the nucleus 
$^{194}$Pt, in Table~\ref{tab:e2-pt} the 3D results 
for the $B(E2;4^{+}_{3}\to I_{f})$ transition rates 
are compared with the experimental 
$B(E2;4^{+}_{2}\to I_{f})$ values.  

An empirical fact for $\gamma$-soft nuclei is that  
the $B(E2;2^{+}_{2}\to 2^{+}_{1})$ value is large and 
of the same order of magnitude as 
the $B(E2;2^{+}_{1}\to 0^{+}_{1})$. 
For the Os isotopes, all three IBM 
calculations considerably overestimate the 
experimental $B(E2;2^{+}_{2}\to 2^{+}_{1})$ values. 
For the Pt isotopes, in turn, the corresponding 
$B(E2;2^{+}_{2}\to 2^{+}_{1})$ values are considerably
smaller than those predicted for 
the Os nuclei. This discrepancy can be related 
to the result for the excitation spectra shown 
in Fig.~\ref{fig:band_all_pk}, 
namely, that the $2^{+}_{2}$ band-head of the $\gamma$-band 
is systematically overestimated in Pt. 
In addition, there is no significant 
difference in the  
$B(E2;2^{+}_{2}\to 2^{+}_{1})$ values predicted by the 
three IBM calculations, because the calculation 
suggests that both the 
ground state and $\gamma$ bands predominantly belong to 
the normal $[n_{0}]$ configuration. 

In the O(6) limit of the IBM, the $0^{+}_{2}$ state 
is interpreted to belong to the same $\sigma=n_{0}$ family as the 
ground-state band, with the O(5) quantum number $\tau=3$. 
Especially for the heavier nuclei considered, e.g., 
$^{192}$Os and $^{196}$Pt, the measured $E2$ transition 
rates show the pattern that is close to the O(6) prediction, 
characterized by the large 
$B(E2;0^{+}_{2}\to 2^{+}_{2})/B(E2;0^{+}_{2}\to 2^{+}_{1})$ 
ratio. 
As seen from Table~\ref{tab:e2-os} and Table~\ref{tab:e2-pt}, 
the 2D-$\beta\gamma$ calculations for these nuclei 
provide results that exhibit this $E2$ selection rule 
and that are in agreement with data. 
However, by the inclusion of the pairing 
the $B(E2;0^{+}_{2}\to 2^{+}_{2})$ transition 
rates are lowered by two orders of magnitude 
in most of the Os and Pt nuclei, while 
the $B(E2;0^{+}_{2}\to 2^{+}_{1})$ rates are increased, 
leading to the almost vanishing  
$B(E2;0^{+}_{2}\to 2^{+}_{2})/B(E2;0^{+}_{2}\to 2^{+}_{1})$ 
ratio. 
In the 2D-$\alpha\beta$ and 3D-$\alpha\beta\gamma$ 
calculations the $0^{+}_{2}$ 
states are dominated by the pair vibrational 
configurations (see, Fig.~\ref{fig:wf_ex0}) for 
all the nuclei and, consequently, do not follow 
the $E2$ selection rule 
that is expected by the O(6) symmetry. 
It appears, therefore, that the $0^{+}_{2}$ states 
obtained by the present calculations that involve 
the pairing degree of freedom
should not be associated with 
the $0^{+}_{2}$ state in the O(6) limit. 
In fact, the $0^{+}_{3}$ states for $^{192}$Os and $^{196}$Pt 
obtained from the 3D-$\alpha\beta\gamma$ 
as well as 2D-$\alpha\beta$ calculation, which are 
mainly composed of the normal configuration (Fig.~\ref{fig:wf_ex0}), 
in turn, exhibit large 
$B(E2;0^{+}_{3}\to 2^{+}_{2})/B(E2;0^{+}_{3}\to 2^{+}_{1})$ 
ratio, which is expected in the O(6) symmetry for the 
$0^{+}_{2}$ state.

Some $B(E2)$ values can be used as  
quantitative measures 
that differentiate between various limits of dynamical 
symmetries of the IBM and/or of the geometrical 
models of $\gamma$-soft nuclei. 
Figure~\ref{fig:e2} depicts the calculated $B(E2)$ ratios  
(a) 
$R_{2\gamma}\equiv 
B(E2;2^{+}_{\gamma}\to 2^{+}_{g})
/B(E2;2^{+}_{g}\to 0^{+}_{g})$, 
(b) 
$R_{02}\equiv 
B(E2;0^{+}_{2}\to 2^{+}_{g})
/B(E2;2^{+}_{g}\to 0^{+}_{g})$, 
and 
(c) 
$R_{3\gamma}\equiv 
B(E2;3^{+}_{\gamma}\to 2^{+}_{g})
/B(E2;2^{+}_{g}\to 0^{+}_{g})$
for Os and Pt nuclei. 
In Fig.~\ref{fig:e2}(a), 
the computed ratios $R_{2\gamma}$ from all
the three IBM calculations are close to the 
$\gamma$-unstable O(6) limit $R_{2\gamma}=10/7$, 
both for Os and Pt. One notices that, in fact, for the Os nuclei this 
is at variance with the data, 
which are closer to the rotational SU(3) limit 
$R_{2\gamma}=0$. 
This can be attributed to the fact 
that the SCMF-PESs suggest 
pronounced $\gamma$-softness in the Os chain. 
The values of the calculated ratio $R_{02}$, 
depicted in Fig.~\ref{fig:e2}(b), 
are relatively small $<0.1$.
In particular, this is the case for the 
2D-$\beta\gamma$ IBM 
results that are close to the experimental values 
and the O(6) limit $R_{02}=0$. 
The inclusion of the pairing 
degree of freedom does not seem to improve 
the description of this quantity. 
The ratio $R_{3\gamma}$ differentiates 
between the rigid-triaxial-rotor 
and $\gamma$-unstable-rotor 
(equivalent to O(6) limit of the IBM) 
limits. With the restriction to axial symmetry in the 
2D-$\alpha\beta$ calculations, this ratio 
is below the Wilets-Jean limit of 
$R_{3\gamma}=1.19$. The inclusion of 
triaxiality in the 2D-$\beta\gamma$ and 
3D-$\alpha\beta\gamma$ IBM calculations, 
leads to an increase of the $R_{3\gamma}$ ratio, 
such that it lies between the two geometrical limits.

%-----------------------------------------------------------------------
%
%	\rhoE0
%
%-----------------------------------------------------------------------
\begin{table}[!htb]
\begin{center}
\caption{\label{tab:rhoe0} 
Comparison between the 
experimental \cite{wood1999} and theoretical 
$\rho^{2}(E0;I_{i}\to I_{f})\times 10^{3}$ values. 
The theoretical values are obtained from 
IBM calculations that include triaxial quadrupole 
(denoted by $\beta\gamma$), 
axial plus dynamical pairing ($\alpha\beta$), 
and triaxial plus dynamical pairing ($\alpha\beta\gamma$) 
degrees of freedom. The IBM calculations are based on the 
PC-PK1 energy density functional.
}
 \begin{ruledtabular}
 \begin{tabular}{llcccc}
  & $I_{i}\to I_{f}$ & EXP 
& $\beta\gamma$ & $\alpha\beta$ & $\alpha\beta\gamma$ \\ 
\hline
$^{188}$Os 
& $0^+_2\to 0^+_1$ & 0.013$\pm 0.005$ & 0.00027 & 0.30 & 0.013 \\ % Wood 1999
& $2^+_2\to 2^+_1$ & 0.7$\pm 0.6$ & 0.027 & 0.48 & 0.056 \\ % Wood 1999
$^{194}$Pt
& $0^+_2\to 0^+_1$ & 0.16$\pm 0.08$ & 0.0052 & 0.45 & 0.16 \\ % Wood 1999
& $0^+_4\to 0^+_1$ & 11$\pm 4$ & 47 & 42 & 36 \\ % Wood 1999
& $2^+_2\to 2^+_1$ & 0.46$\pm 0.16$ & 0.011 & 0.036 & 0.018 \\ % Wood 1999
$^{196}$Pt
& $0^+_2\to 0^+_1$ & $<0.07$ & 0.34 & 0.47 & 0.31 \\ % Wood 1999
& $0^+_3\to 0^+_1$ & $<18$ & 17 & 0.61 & 0.61 \\ % Wood 1999
& $0^+_3\to 0^+_2$ & $<39$ & 0.069 & 0.0048 & 0.00025 \\ % Wood 1999
& $2^+_2\to 2^+_1$ & 1.0$\pm 0.6$ & 0.022 & 0.025 & 0.019 \\ % Wood 1999
 \end{tabular}
 \end{ruledtabular}
\end{center} 
\end{table}
%-----------------------------------------------------------
%       E0, PC-PK1
%-----------------------------------------------------------
\begin{figure}[ht!]
\begin{center}
\includegraphics[width=\linewidth]{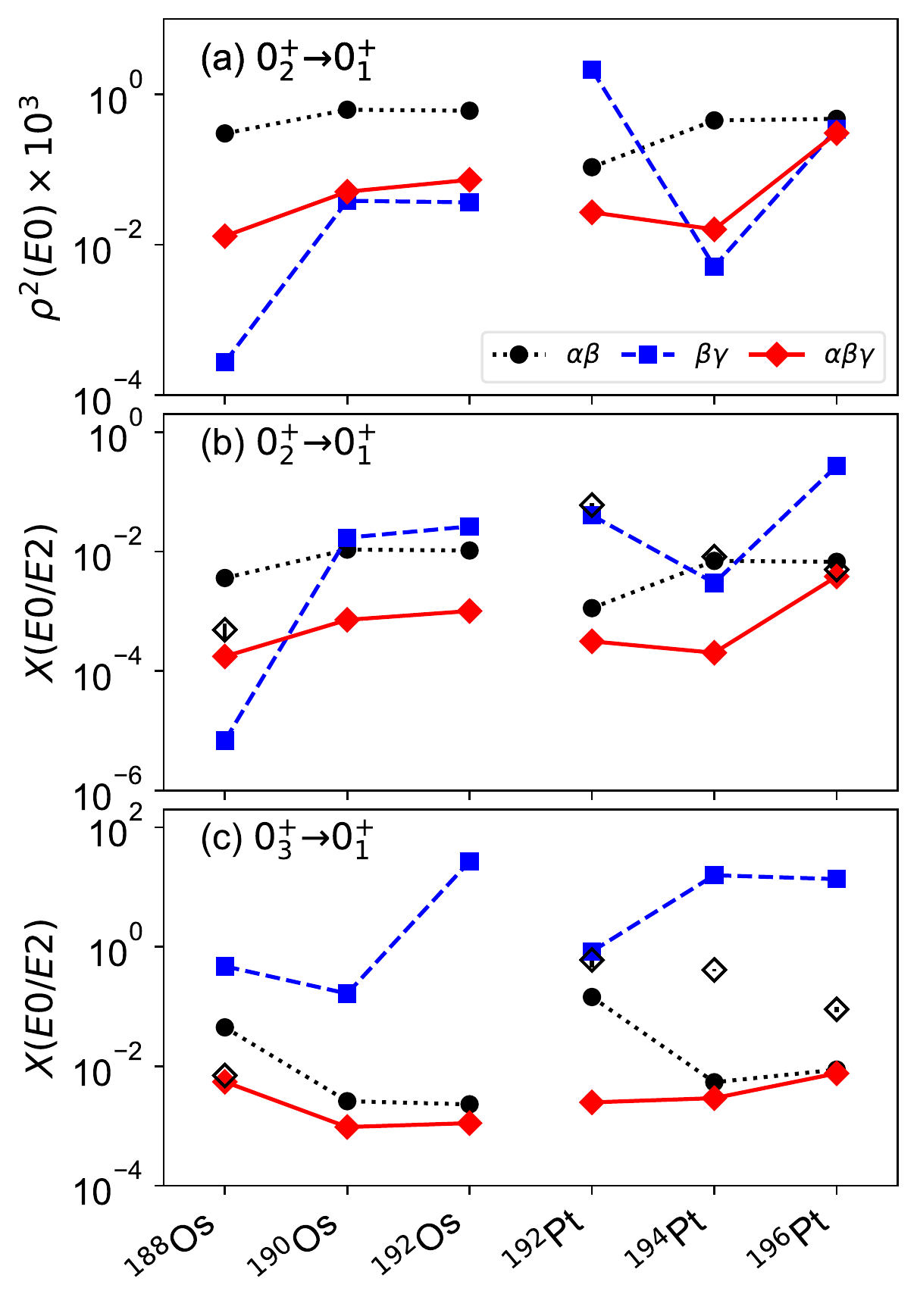}
\caption{The 
$\rho^{2}(E0; 0^{+}_{2}\to 0^{+}_{1})$ values 
and $X(E0/E2)$ ratios for the 
$0^{+}_{2}\to 0^{+}_{1}$ and 
$0^{+}_{3}\to 0^{+}_{1}$ transitions, 
for $^{188,190,182}$Os and $^{192,194,196}$Pt.
The PC-PK1 functional is used in the IBM calculations.
The experimental values are taken from \cite{data,kibedi2005} 
for the $X(E0/E2)$ ratios, and are 
represented by the open symbols.} 
\label{fig:e0}
\end{center}
\end{figure}

\subsubsection{$E0$ transition properties\label{sec:e0}}
Table~\ref{tab:rhoe0} lists the 
$\rho^{2}(E0)$ values for $^{188}$Os, $^{194}$Pt, 
and $^{196}$Pt, for which limited experimental 
results are available. The theoretical values are the results of 
IBM calculations including triaxial quadrupole 
($\beta\gamma$), 
axial plus dynamical pairing ($\alpha\beta$), 
and triaxial plus dynamical pairing ($\alpha\beta\gamma$) 
degrees of freedom. 
As one notices from the 2D-$\alpha\beta$, and 
the 3D-$\alpha\beta\gamma$ results, the inclusion of triaxiality 
generally decreases  the $\rho^{2}(E0)$ values. 
The 2D-$\alpha\beta$ results are in better agreement 
with the experimental values, compared to the $(\beta,\gamma)$ ones. 
Thus the pairing degree of freedom appears to be more important than 
triaxiality in describing $E0$ transitions. 
However, because of a complex interplay between 
both degrees of freedom and also due 
to the presence of adjustable parameters in the 
$E0$ operator, it is 
not straightforward to draw a generic conclusion 
about the relevance of considering both 
triaxial and pairing deformations 
in the calculation of the $E0$ properties. 
For similar reasons, and also because of the 
present assignment of the $0^{+}_{2}$ states to be mainly 
of pair-vibrational nature, 
the $\rho^{2}(E0;0^{+}_{2}\to 0^{+}_{1})$ value for $^{196}$Pt, 
which should vanish in the O(6) limit, is calculated 
to be much larger than the upper limit of the 
corresponding experimental value. 

Figure~\ref{fig:e0} displays  
the $\rho^{2}(E0; 0^{+}_{2}\to 0^{+}_{1})\times 10^{3}$ 
values and the $X(E0/E2)$ ratios for the 
$0^{+}_{2}\to 0^{+}_{1}$ and 
$0^{+}_{3}\to 0^{+}_{1}$ transitions. 
The mixing ratio $X(E0/E2)$ reads
\begin{align}
 X(E0/E2)
=\frac{\rho^{2}(E0;0_{i}^{+}\to 0^{+}_{f})e^{2}R^{4}}
{B(E2;0^{+}_{i}\to 2^{+}_{1})}
\end{align} 
with $R=1.2A^{1/3}$ fm. 
As shown in Fig.~\ref{fig:e0}, the considered $E0$ 
transition properties are quite sensitive to the 
nature of the $0^{+}$ states, and hence can differ  
by orders of magnitude between neighboring isotopes. 
For the $X(E0/E2)$ ratios, in particular, 
the $B(E2;0^{+}_{i}\to 2^{+}_{1})$ value in the denominator 
is, in some cases, negligible, resulting 
in an unusually large mixing ratio. 
The simultaneous inclusion of triaxial and pairing 
deformations tends to result in a   
$\rho^{2}(E0; 0^{+}_{2}\to 0^{+}_{1})$ value 
that is the smallest among the three types of 
IBM calculations, except for $^{192}$Pt. 
From both Figs.~\ref{fig:e0}(b) and \ref{fig:e0}(c), 
one notices that the $X(E0/E2)$ results obtained with the 
3D-$\alpha\beta\gamma$ IBM, are the 
smallest among the three different calculations. 
The same trend is observed in the $X(E0/E2)$ 
results obtained with the DD-PC1 EDF.

\section{Conclusions\label{sec:conclusion}}
Based on the framework of nuclear EDFs, 
the effects of coupling between quadrupole triaxial 
shape and dynamical pairing degrees of freedom 
have been investigated
in spectroscopic calculations of low-energy collective states  
of $\gamma$-soft nuclei. 
Constrained SCMF calculations have been 
performed using the RMF+BCS method with 
a choice of universal EDF and pairing 
interaction, resulting in potential energy 
surfaces as functions of the triaxial quadrupole 
$(\beta,\gamma)$ and pairing $\alpha$ degrees 
of freedom (the coordinate $\alpha$ is proportional to the pairing gap $\Delta$) 
for typical $\gamma$-soft nuclei 
in the mass $A\approx 130$ ($^{128,130}$Xe) 
and $A\approx190$ 
($^{188,190,192}$Os and $^{192,194,196}$Pt) regions.

The SCMF deformation energy surfaces for all considered nuclei 
exhibit notable softness in $\alpha$ and $\gamma$, 
thus pointing to the importance of 
correlations that arise from fluctuations of triaxial 
and pairing deformations. 
Spectroscopic properties have been  
computed by employing the 
boson-number non-conserving IBM Hamiltonian, consisting 
of up to three-body boson terms. 
The parameters of the IBM Hamiltonian 
have been determined by mapping the 
SCMF-PES onto the expectation value 
of the Hamiltonian in the boson condensate state. 
The mapped IBM framework that simultaneously 
takes into account the dynamical pairing 
and quadrupole triaxial degrees of freedom has 
shown that: (i) the inclusion of 
dynamical pairing significantly lowers 
the energies of the excited $0^{+}$ states and 
structures built on them,  
in good agreement with experimental results;
(ii) the description 
of $\gamma$-vibrational bands and the 
related $B(E2)$ rates is considerably improved 
by the effect of triaxiality; and  
(iii) the principal results are not particularly  
sensitive to the choice of the microscopic EDF.

The results of the present work, 
together with those of the 
exploratory study for $^{128,130}$Xe 
\cite{nomura2021pv}, 
clearly demonstrate the importance of 
simultaneously including the dynamical pairing and 
quadrupole triaxial shape degrees of freedom, 
and their explicit coupling, 
for a quantitative description of 
low-energy collective states of medium-mass 
and heavy nuclei. 
The method developed here and in the previous work, 
can be used to explore interesting structure 
phenomena, such as shape phase transitions and 
shape coexistence in $\gamma$-soft and triaxial nuclei, 
in which cases both pairing vibrations 
and triaxial deformations are expected 
to play a significant role.

\begin{acknowledgements}
This work has been supported by the 
Tenure Track Pilot Programme of 
the Croatian Science Foundation and the 
\'Ecole Polytechnique F\'ed\'erale de Lausanne, 
and the Project TTP-2018-07-3554 
Exotic Nuclear Structure and Dynamics, 
with funds of the Croatian-Swiss Research Programme. 
It has also been supported in part by the QuantiXLie Centre of Excellence, a project co-financed by the Croatian Government and European Union through the European Regional Development Fund - the Competitiveness and Cohesion Operational Programme (KK.01.1.1.01.0004).
The author Z.P.L. acknowledges support by the NSFC under Grant
No. 11875225. The author J.X. acknowledges support by the NSFC under
Grants No. 12005109 and No. 11765015. 
\end{acknowledgements}

\bibliography{refs}

\end{document}